\documentclass[nonacm, screen, sigconf]{acmart}

\usepackage{tikz}
\usepackage{cleveref}
\usepackage{pstricks}
\definecolor{links}{RGB}{30, 85, 255}
\definecolor{cites}{RGB}{30, 200, 30}
\definecolor{urls}{RGB}{255, 116, 20}

\newcommand{\pie}[1]{%
\begin{tikzpicture}
 \draw (0,0) circle (1ex);\fill (1ex,0) arc (0:#1:1ex) -- (0,0) -- cycle;
\end{tikzpicture}%
}

\def\showauthnotes{1}
\ifthenelse{\showauthnotes=1}
{
	\newcommand{\authnote}[2]{{ \footnotesize \bf{[#1: #2]~}}}
}
{ \newcommand{\authnote}[2]{} }

\newtheorem{definition}{Definition}
\newtheorem*{remark*}{Remark}

\newcounter{subdefinition}[definition]
\renewcommand{\thesubdefinition}{\thedefinition.\arabic{subdefinition}}

\AtBeginDocument{%
  \providecommand\BibTeX{{%
    \normalfont B\kern-0.5em{\scshape i\kern-0.25em b}\kern-0.8em\TeX}}}

\begin{document}
\hypersetup{linkcolor={links},citecolor=[named]{cites},urlcolor={urls}}

\title{SoK: Blockchain Governance}



\author{Aggelos Kiayias}
\affiliation{%
  \institution{University of Edinburgh, IOHK}
  \country{United Kingdom}
}
\email{aggelos.kiayias@ed.ac.uk}

\author{Philip Lazos}
\affiliation{%
  \institution{IOHK}
  \country{United Kingdom}
}
\email{philip.lazos@iohk.io}

\renewcommand{\shortauthors}{}

\begin{abstract}
Blockchain systems come with a promise of decentralization that, more often than not, stumbles on a roadblock when key decisions about modifying the software codebase need to be made. In a setting where ``code-is-law,'' modifying the code can be a controversial process, frustrating to system stakeholders, and, most crucially, highly disruptive for the underlying systems. This is attested by the fact that both of the two major cryptocurrencies, Bitcoin and Ethereum, have undergone ``hard forks'' that resulted in the creation of alternative systems which divided engineering teams, computational resources, and duplicated digital assets creating confusion for the wider community and opportunities for fraudulent activities. The above events, and numerous other similar ones, underscore the importance of Blockchain governance, namely the set of processes that blockchain platforms utilize in order to perform decision-making and converge to a widely accepted direction for the system to evolve. While a rich topic of study in other areas, including social choice theory and electronic voting for public office elections, governance of blockchain platforms is lacking a  well established set of methods and practices that are adopted industry wide. Instead, different systems adopt approaches of a variable level of sophistication and degree of integration within the platform and its functionality. This makes the topic of blockchain governance a fertile domain for a thorough systematization that we undertake in this work. 

Our methodology starts by distilling a comprehensive array of properties for sound governance systems drawn from academic sources as well as grey literature of election systems and blockchain white papers.  These are divided into seven categories, suffrage, Pareto efficiency, confidentiality, verifiability, accountability, sustainability and liveness that capture  the whole spectrum of desiderata of governance systems. We interpret these properties in the context of blockchain platforms and proceed to classify ten blockchain systems whose governance processes are sufficiently well documented in system white papers, or it can be inferred by publicly available information and software. While all the identified properties are satisfied, even partially, by at least one system, we observe that there exists no system that satisfies most properties. Our work lays out a common foundation for assessing governance processes in blockchain systems and while it highlights shortcomings and deficiencies in currently deployed systems,  it can also be a catalyst for improving these processes to the  highest possible standard with appropriate trade-offs, something direly needed for blockchain platforms to operate effectively in the long term. 
\end{abstract}




\maketitle

\section{Introduction}
Following the founding of Bitcoin \cite{nakamoto2008bitcoin} in 2009, cryptocurrencies and other blockchain platforms have tremendously risen in popularity. Unlike centralised organisations, which are governed by a select few, blockchain platforms operate in a decentralised fashion by the different actors in these platforms. The decentralised nature of blockchains has been essential to their appeal; however, it has also introduced new challenges. Blockchain platforms, like other organisations, try to adapt and adjust to their stakeholders’ needs and preferences. With different actors present whose preferences might not always align, governance problems arise and the risk of division between their community members increases.

Different governing mechanisms exist, depending on the platform. Off-chain governance is the most centralised of such mechanisms with the core developers or the most trusted contributors making most of the decisions. On-chain governance is achieved via on-chain voting mechanisms, which can be more transparent and inclusive than off-chain governance. In both of these mechanisms, community division can take place when a backward-incompatible update is adopted, where some stakeholders choose to stay on the original chain and others choose to upgrade to the updated chain, dividing the community into two. Alternatively, two or more competing updates may be proposed dividing the community about their potential merits. Eventually, consensus can fail and different segments of the community  adopt the update that they believe to be the most beneficial. 

In the most general sense, such deviations are known as hard forks and numerous examples of them have been observed in popular cryptocurrencies. 
Two notable examples are the split of the Ethereum chain to Etheurem and Ethereum Classic 
due to the the DAO debacle \cite{dao_hack_report}
and the split of the Bitcoin system into Bitcoin and
Bitcoin Cash over the debate around block size and the SegWit upgrade. 
Such divisions can fragment the community and its resources, and as a result reduce the overall value of the platform as well as its security. The latter consideration can be quite tangible as the reduced number of resources supporting a fork can lead to attacks. Such attacks are referred to as  $51\%$ attacks and have occurred on a number of occasions, e.g., see the case of Ethereum Classic \cite{etc51} for a notable such instance. 

The above issues highlight the importance of  sound blockchain governance, the ability of a blockchain platform community members to express their will effectively regarding the future evolution of the platform as well as the best possible utilization of its resources. So this brings forth the question what characterizes  proper governance in blockchain systems? This fundamental  question motivates the systematization effort we undertake in this paper. 

Our methodology is first to derive a set of properties,
that are drawn from general governance principles and election theory and then interpret them to the  blockchain governance setting. 
We use a variety of sources to ensure the comprehensiveness of our property list that include the Council of Europe technical standards for e-voting \cite{CoE}, the Federal Election  Commission's Voting Systems Standards \cite{FEC},
but also blockchain specific ones such as \cite{vitalik3, systematic_review, wharton}. 
Given the set of properties,  we then evaluate a wide array of blockchain platforms against  those properties revealing each platform's unique strengths and weaknesses. 

We distill seven fundamental properties for blockchain governance. The properties capture different aspects
of important requirements for governance. 
The  first property deals with participation eligibility;
Decision making systems can produce 
legitimate outcomes provided they are inclusive --- a property that we capture by different aspects of {\em Suffrage} suitably adapted to the blockchain setting.
Suffrage determines a set of ``decision-makers'' who are a subset of the community of a blockchain project. 
The second property has to do  with
the {\em Confidentiality} of the decision-makers' inputs; it further specializes to Privacy, which asks for maintaining the input private while Coercion Resistance asks for the input to be free of any external influences. The third property ---{\em Verifiability}--- asks 
for decision-makers to be able to verify their input has been taken into account in the output and that such output is correctly computed. These last two properties are in a sense ``classical'' security properties. Next we move to two properties that have to do with the incentives of the decision-makers. {\em Accountability} asks for decision-makers to be held accountable for the input they provide to the system, while {\em Sustainability} asks whether appropriate incentives are provided for the system to evolve constructively and to the decision-makers for providing meaningful input. We then move to a social choice consideration. {\em Pareto efficiency} asks that, given all decision-makers' preferences, the outcome of the governance process cannot be strictly improved vis-\`a-vis these preferences.  Finally, the crucial ability of the system to produce outputs expediently is captured by {\em Liveness.}

Armed with the above comprehensive list of governance properties we investigate a number of popular blockchain platforms which provide some sort of governance functionality and we detail  the way they satisfy (or fail to satisfy) each of the given properties. Our results dictate that while each of the properties is considered in the context of at least one system, there exists no platform that satisfies most of the properties. 


\subsection{Related Work}
As of the time of writing, there is yet to be a formal or rigorous coverage of good blockchain governance properties. However, the topic of blockchain governance has received coverage in multiple disciplines. Given their diversity, additional related work is also presented in context within each subsection of Section~\ref{section:properties}, where each governance property is defined. \citet{defining_blockchain_governance} adapt the definition of OSS (open-source software) governance to blockchain governance; they then go on to derive six dimensions and three layers of blockchain governance from the literature to build a framework, which can be used as a starting point for discussion in new blockchain projects. Similarily \citet{gov_in_blockchain_economy} derive three key dimensions of blockchain governance to define an IT governance definition. \citet{gov_of_and_by_the_infra} investigate the social and technical governance of Bitcoin, making a distinction between two coordination mechanisms: governance by the infrastructure (via the protocol) and governance of the infrastructure (by the community of developers and other stakeholders). Corporate governance has been drawn from in the literature to examine the governance of public blockchains. The work done by \citet{internal_external_governance} and \citet{endogenous_exogenous_governance} are such examples, where the authors of the latter work derive a definition of blockchain governance and make a distinction between endogenous and exogenous governance. Given the variety of actors and strategies in the decision-making processes in blockchain platforms, \citet{blockchain_nash_equillibria} view blockchain governance from the lens of IT governance and then analyse decision-making processes in the form of voting on a new blockchain improvement proposal, by using Nash equilibria to predict optimal governance strategies. Certain forms of blockchain governance, like traditional forms of governance, have the short-coming of participants not able to change their vote between two consecutive elections or votes. \citet{always_on_voting} address this shortcoming, among others, by introducing an always-on-voting (AoV): a repetitive blockchain-based voting framework that allows participants to continuously vote and change elected candidates or policies without having to wait for the next election. More specific analysis on certain aspects of blockchain decision-making processes also exist in the literature (e.g. \citet{delegation_misbehaviour} where the authors analyse delegated voting and conclude caution should be exercised when implementing such mechanisms).

\section{Blockchain Governance Properties}\label{section:properties}

One of the main contributions of our work is systematizing the properties pertinent to blockchain governance systems. 
We would like to stress that there is no \emph{single set} that optimally captures every aspect. There are  trade-offs between satisfying some properties to a high degree and others to a lesser degree. In addition, many current implementations do not have rigorously defined governance mechanisms for every use case and usually contain a mixture of formal on-chain features as well as informal off-chain ones. This is almost inevitable, as different blockchains are built for specific purposes and not all decision-making processes can be  sufficiently captured  by a smart contract or special purpose protocol logic. Others might still be centralized or transitioning to full decentralization. 
Irrespective of this, our property systematization focuses on {\em first principles} and is meaningful across the board, independently of the underlying set of mechanisms that are set in place to facilitate decision-making in each blockchain platform. 

\begin{figure}[h]
    \centering
    \includegraphics[scale=0.25]{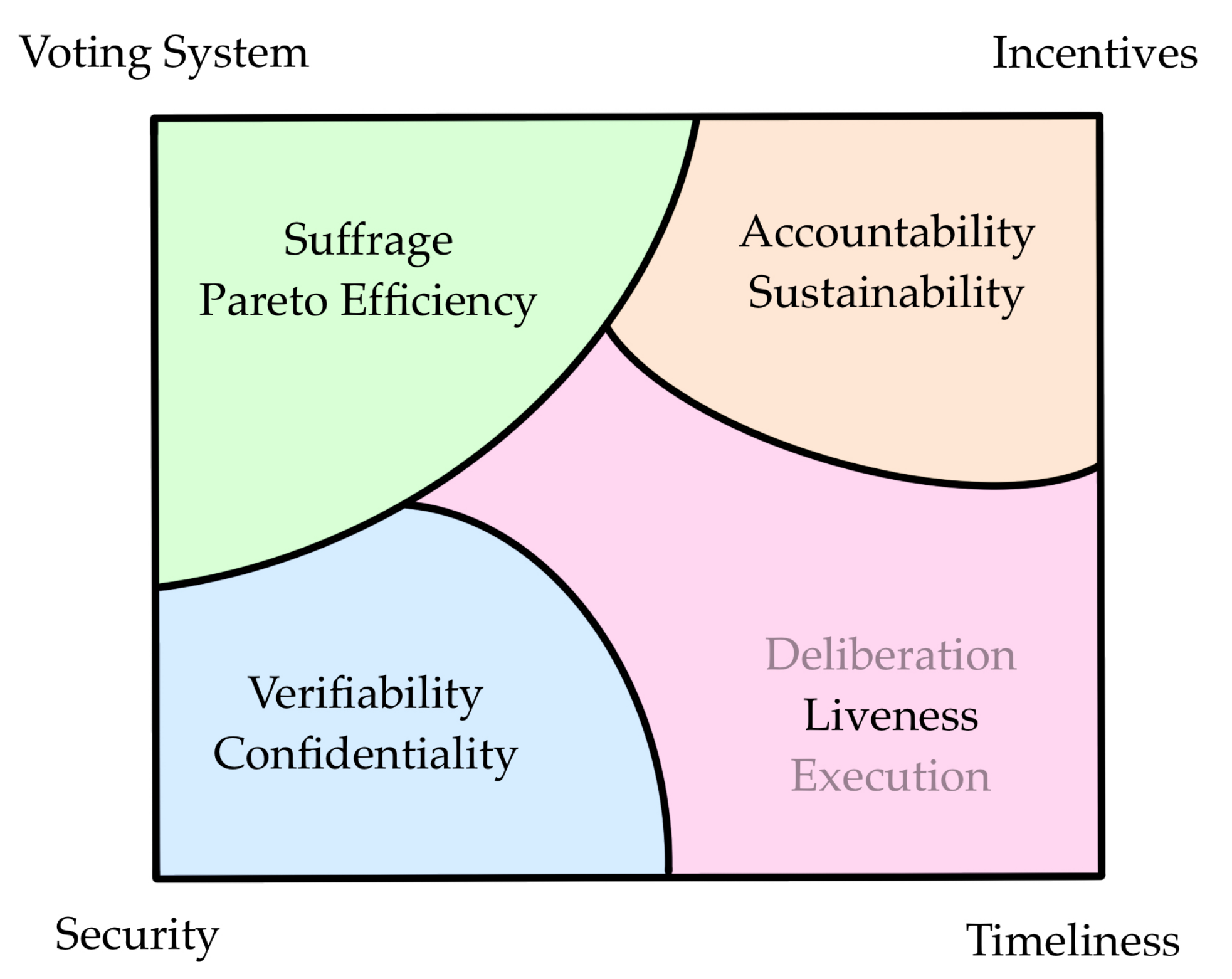}
    \caption{The partition map of governance properties.}
    \label{fig:platonic}
\end{figure}

We can categorize the properties into four broad classes pictorially shown in \Cref{fig:platonic}. The first class contains properties about the {\em voting system } 
that is used for decision-making. It will touch the issues of who is eligible to participate and what is the process that combines the inputs provided. 
The voting system enables us to argue about the governance process in an ideal, philosophical sense; questions such as who has the right to vote are relevant here. 
The remaining three classes deal with the way an ideal voting system can be implemented 
and touch three important domains: {\em security} which deals with cryptographic and cyber-security aspects,  {\em incentives} which deals with game-theoretic and economics aspects, and {\em timeliness} which deals with issues of time and expediency.  Within, the keywords \emph{Deliberation} and \emph{Execution} are greyed-out. These   are not the focus of   our systematization. The reasoning behind this will be explained below.  
Failures in the properties of these classes can have important repercussions 
for the legitimacy of the governance process. Even though the voting system 
might be acceptable in a `Platonic' ideal sense, failures in the remaining properties
can suggest that certain community members are disenfranchised because it is harder
for them to participate, or they cannot express their will freely or even that
they have no ability to properly form an opinion due to lack of proper incentivization. 
%
%
It is also worth adding  that {\em usability} permeates these three implementation related classes, but it will be outside of scope of our systematization. 


An important aspect of our property systematization is that we emphasize fundamental properties entirely decoupling them from any specific techniques, algorithms or mechanisms that facilitate them. 
To illustrate the point, a simple example is the distinction between the property of having privacy (or secrecy) and the cryptographic protocol techniques that may be used to achieve it. Another example is quadratic voting, which is a technique where additional votes can be `bought' (using actual money, voting credit, etc.) but the cost scales quadratically with the number of votes. Even though 
it has received renewed interest in blockchain governance, particularly for participatory budgeting applications,\footnote{Such as Gitcoin quadratic funding,  \url{https://gitcoin.co/blog/gitcoin-grants-quadratic-funding-for-the-world/}} it should be clear it is still just a \emph{mechanism}, not a fundamental property per se; we revisit it in some more detail when we discuss Suffrage below as it is one of our  basic properties that is most related.

Further to this point, whether a particular governance mechanism is on-chain, off-chain, uses a foundation etc. is a \emph{mechanism}, \textbf{not} a property. These inner workings will not be part of our classification explicitly, unless they affect some fundamental property.

We want to stress that satisfying all properties to some higher or lower degree, as feasible, would not make a blockchain governance system perfect. There are many blockchains applications and each of them has different needs and use cases that would require community involvement. Some properties might be incompatible with each other. Our thesis though is that any design would have to \emph{consider} how each property is addressed and ensure that the the choices made are deliberate. As such, during the evaluation of different governance systems we will make sure that each property is judged \emph{in context}, taking the goals of each system into account.

\paragraph{Timeliness: scope and limitations}
In describing the  properties (excluding, to an extent, \emph{liveness}) we take a high-level, theoretical approach, obviating the need to explain the underlying social and organizational structures that power the governance systems which exhibit them. This is intentional: by remaining abstract we can cover a sufficiently large `space' of governance system designs, without sacrificing too much detail. 
%
Still, it  is important 
to acknowledge such structures as they are an integral part of governance. 
First, it should 
be possible (and easy) for the users to \emph{deliberate} (often done through github, Discord or public internet forum) in order to converge on the topics that need to enter the decision-making process. Second, the manner which the executive power is conferred is critical. In some (simpler) cases it is possible to make execution automated --- while other times larger structures (such as a private enterprise or non-profit foundation) can engage to implement the outcomes of the decision making process. As a takeaway, our focus will be the study of governance process in between Deliberation and Execution, assuming both of them are feasible.

\subsection{Suffrage}\label{subsec:suffrage}
One of the first considerations of any governance system is determining who is granted \emph{suffrage}, which is the right to participate in decision making procedures. This can be distinguished in \emph{active} suffrage, the right to vote, and \emph{passive} suffrage, which is the right to stand for election and become an elected representative.
Suffrage, an already a complicated and nuanced property, is even more so when applied to blockchain systems. 

\newcommand{\citizen}{C}
\newcommand{\decisionmaker}{D}
In national or regional elections, it is often the case that the voting mechanism implements a `one person, one vote' rule. Different jurisdictions use different criteria in guaranteeing the right to vote to individuals, but the bottom line is that one person can only submit one vote. Although research is currently underway on proof-of-personhood systems \cite{proof_of_personhood}, which verify that accounts correspond to unique individuals, the `one person, one vote' rule is not applicable to most, if not any, current blockchain platforms.  Instead, we often see that a minimum amount of stake or hashing power is required to guarantee a vote. We also see platforms where only the founders or core developers are guaranteed a vote. In any case, these are attempts to define and reconcile two groups of people: the set of community-members $\citizen$ and decision-makers $\decisionmaker$.
\begin{definition}
The community-members $\citizen$ of a blockchain system are people that have \emph{direct} interaction with it. This may be by providing resources in service of its security or consensus protocol, owning tokens, develop software etc.
\end{definition}
\begin{definition}\label{def:decisionmaker}
The decision-makers $\decisionmaker \subseteq \citizen$ of a blockchain system are the people that participate in (any way) its governance.
\end{definition}
 
Given these definitions, we establish the basic ways that com\-munity-members are granted voting rights in the blockchain space. The voting rights should more accurately be called voting \emph{weights}, as it is very common to allocate a different number of votes across all community-members.
\begin{definition}[Type 1: Identity-Based Suffrage]
\textit{A blockchain governance system satisfies this property if it guarantees decision-making rights to participants who are able to prove their identities such that the votes correspond to unique individual humans}.
\end{definition}
Contrary to the usual notion of community-membership, identity alone is not (so far) a robust enough connection between users and blockchains. Also, there is no restriction against switching to different blockchains or having direct interactions with many of them. The following notions of suffrage are based on a more `quantifiable' approach and typically assign voting power accordingly.
\begin{definition}[Type 2: Token-Based Suffrage]
\textit{A blockchain governance system satisfies this property if it guarantees decision-making rights to participants who have certain tokens in the platform or a minimum amount of tokens in the platform}.
\end{definition}

\begin{definition}[Type 3: Mining-Based Suffrage]
\textit{A blockchain governance system satisfies this property if it guarantees decision-making rights to participants who have a certain amount of hashing power in the platform (or other physical resource relevant to the platform, e.g., disk storage)}.
\end{definition}

In the PoS setting, voting weight is often measured by an operator's stake (or wealth). This can result in the following undesirable situations: \textbf{(i)} participants who may be more enthusiastic about the platform have lower voting weight than those who are less enthusiastic about the platform, and \textbf{(ii)} participants who may have contributed more to the platform may have lower voting weight than those who contributed less. Methods like quadratic voting \cite{quadratic_voting} can help dampen the effects of stake-based voting weight (see below for an explanation), but it does not address the root of the problem: voting weight is ultimately based on wealth owned or even managed (e.g., centralized cryptocurrency exchanges may control a significant amount of stake that does not belong to them). 
Similar issues exist in the PoW setting, where hashing power may not proportionately reflect stakeholder contributions to the platform. 
Analysis in quantifying decentralisation \cite{quantifying_decentralisation} on blockchain platforms, in terms of stake and hashing power, can provide insights into resultant power concentrations. 

\begin{remark*}[Governance Tokens]
Often, tokens used to determine suffrage can have more than one use (e.g., native currency of a proof-of-stake system). However, particularly for the governance of smart contract based protocols, specific \emph{governance} tokens can be used, who have no other direct functionality or value (such as paying for transaction fees or appearing as block rewards) other than enabling participation. Especially when these tokens are transferable, special care is needed to ensure that their supply, distribution and price accurately represents the community members who are more invested in the project. This was observed in the recent Beanstalk \href{https://bean.money/blog/beanstalk-governance-exploit}{exploit}, where an attacker used a flash loan to obtain a majority of governance tokens, passing his own malicious proposal and quickly implementing it. The voting mechanism worked well: but clearly, the voting weights did not accurately reflect the community. To avoid such attacks, other platforms such as Compound employ more fail-safes, such as a mandatory waiting period before enacting the election result.
\end{remark*}

Instead of assuming that community-members would have an implied incentive to positively contribute to their respective block\-chain's governance, sometimes a more direct approach is taken. Participants are granted a decision-making right based on whether they have positively contributed to the platform. What defines a `positive' contribution is not always clear cut and its definition is left to the platform's community.

\begin{definition}[Type 4: Meritocratic Suffrage]
\textit{A blockchain governance system satisfies this property if it only guarantees decision-making rights to participants who have positively contributed to the platform}. 
\end{definition}

\begin{definition}[Type 5: Universal Suffrage]
\textit{A blockchain governance system satisfies this property if it guarantees decision-making rights to participants who have mining power or tokens in the platform as well as participants with positive contributions to the platform}. 
\end{definition}

We reiterate that it is not our objective to outline specific mechanisms for translating community-membership to voting power. For example, we are not suggesting that an actor's voting weight should be more influenced by previous contributions than by an actor's stake in the platform. Instead, we are suggesting that it is important that all forms of investments and contributions of a community-member (which can be very different across different blockchains) should be considered when formulating voting weight.

In this context, a mechanism that has gained traction recently in the blockchain  context is quadratic voting. In this mechanism,  1 vote would cost 1, but 2 votes would cost 4 and so on.  
Such a mechanism could achieve a  better balance between what  \emph{Token-Based Suffrage} and \emph{Identity-Based Suffrage}: having additional currency within the system does entail enhanced voting rights, but some balancing effect vis-\`a-vis  the one-person one-vote rule seems appropriate.
It also provides a more flexible way of expressing voter preferences. To see this, suppose that, in a governance system where votes can be exchanged for tokens, two voters believe that one vote in favour of some proposal is worth 5 and 10 respectively. By this, we mean that the voters believe investing 1 coin for a vote, would yield a return on investment of 4 and 9 respectively. In the final election, if the first voter is richer they could purchase 100 votes, while the second only buys 3. This would signal that the first voter is particularly in favour of this proposal, but in fact they bought more votes just because they had a higher budget to spare. With quadratic voting, the first voter would acquire 2 votes: the next vote would cost 4, which is not seen as a profitable investment.

\subsection{Pareto Efficiency}\label{sec:Pareto}
Any blockchain governance system will necessarily depend on a number of decision-making procedures: individual, competing preferences have to be collected and combined into specific actions. In this section we try to formalize how well the tools provided by blockchain allow the \emph{decision-makers} (recall \Cref{def:decisionmaker}) to reach their most favourable outcome. Ideally, the result would the same as one chosen by an omniscient algorithm that has collected all their private thoughts and magically chose the `perfect' outcome. As we will see, even the notion of a `perfect' outcome is hard to define (and under most definitions, does not always exist). We stress that this might be \emph{terrible} for the community-members of the blockchain; in this section we only focus on how well the intentions of the decision-makers can be turned into actions. Aligning the intentions of the community-members and decision-makers is a question of suffrage (as well as \emph{Accountability}, which we define in \Cref{sec:accountability}).

The investigation of such decion-making processes is the focus of Social Choice Theory~\cite{brandt2012computational}, which is an entire field of study dedicated to them. One of its crowning early achievements is the famous Arrow's Impossibility Theorem (\citet{arrow1950difficulty}), on voting systems where participants \emph{rank} the possible candidates. Specifically, given a set of alternatives $A = \{a_1, a_2, \ldots, a_n\}$, each voter $i$ submits an ordered vector of the form $a_{i_1} \succ a_{i_2} \succ \ldots \succ a_{i_n}$. Combining the votes should lead to an outcome preference ordering $a_{j_1} \succ a_{j_2} \succ \ldots \succ a_{j_n}$ of the candidates that best represents the voters. Unfortunately Arrow's Theorem states that the following natural properties cannot be satisfied at the same time:
\begin{itemize}
    \item If every voter prefers candidate X over Y, then X is ranked higher than Y in the final outcome. This property is often called \emph{unanimity}.
    \item The order of X and Y in the final outcome depends only on the ordering of X and Y in each voters preference, irrespective of how all other candidates are ordered. This is called \emph{independence of irrelevant alternatives}.
    \item There is no voter who has dictatorial control over the final outcome.
\end{itemize}
Variations of this result have been adapted in many voting settings, even in cases where the voting process does not have to reveal an entire ordering of outcomes (but only to select the `best' one) or when voters have \emph{cardinal} preferences (i.e. they can assign numerical preference values to each candidate). Note that almost all popular voting schemes (such as \emph{approval voting}, where each voter selects a set of acceptable candidates) fall under these definitions. Perhaps the most famous of those impossibility results is the Gibbard-Satterthwaite Theorem (\citet{gibbard1973manipulation, satterthwaite1975strategy}), roughly stating that any voting scenario with more than two candidates is either dictatorial, or subject to \emph{strategic voting} (i.e., voters swaying the outcome by misreporting their actual preferences.

To deal with these impossibilities, the voting procedures used in practice are not required to be optimal in every scenario, but to satisfy certain weaker properties depending on the setting. One such mild property is \emph{Pareto efficiency} (e.g., \cite{rivest2010optimal, kluiving2020analysing}). These properties are tested assuming every voter truthfully reports their preferences.
\begin{definition}
A blockchain governance system is Pareto efficient if whenever a decision-making process is held, alternative X cannot win if there exists another alternative Y that is preferred by at least one participant and no participant prefers X over Y.
\end{definition}
A Pareto efficient governance system would never lead to an outcome that is \emph{clearly} worse than another possible outcome. This property should typically be satisfied (at least when interpreted loosely, as some blockchain systems do not have an entirely rigorous governance model), unless there is good reason not to. 
Evaluating whether this property is satisfied can be tricky because a blockchain governance system contains many interacting components, with the final result seldom depending on a single vote. We make our best effort to fairly evaluate how \emph{likely} it is that a Pareto efficient outcome is not selected and \emph{how} much worse is the selected alternative. 

\emph{Approval voting} is of particular importance, as it is the most common voting mechanism used by the blockchains we evaluate. Given $n$ candidates, each voter can `approve' as many as they want. The winner is the candidate which was approved by most voters, often combined with a threshold, such as also requiring approval from at least $20\%$ of them. Notice that even though the voters might have ordinal or cardinal preferences, they can only submit a binary signal for each candidate. Starting with a simple example, suppose that 2 possible \emph{incompatible} blockchain updates $a$ and $b$ are up for election. Furthermore, suppose that \emph{every} voter prefers $a \succ b$. The outcome will be dictated by the threshold they chose when \emph{converting} their ordinal preferences to an approval vote. Typically we would expect $a$ to win, but $b$ could win as well! Clearly, any truthful voter who approved $b$ would also approve $a$, since $a \succ b$ for every voter. However, some voters might chose \emph{not} to approve either of them. In this case $b$ could win because of a tie. In fact, this is the only way an outcome of approval voting might not be Pareto efficient: if the winner is tied with the Pareto optimal candidate. This happened because the voters where completely uniformed about the preferences of each other and set their `approval threshold' too high. The more information they have the less likely such an outcome becomes. A group of perfectly rational and informed voters would always produce a Pareto efficient outcome.
In addition, it is important to keep in mind that there are two more `secret' (implicit) options always available: to do \emph{nothing} or to \emph{fork}, which is to be avoided. When combined with a minimum approval threshold and some awareness on the part of the voters, the winner is most likely either Pareto efficient, a suboptimal yet highly popular alternative or a deadlock. Finally, strategic voting involves setting the threshold very high, which decreases the total number of votes and could lead to a deadlock, but is unlikely to result in a fork.

We briefly discuss an alternative voting system that uses the complete \emph{ordinal} preference profile called \emph{instant-runoff} (IRV) voting. It proceeds in turns:
\begin{itemize}
    \item From every ballot, only the top preference is counted.
    \item If one candidate obtains a majority, they win.
    \item Otherwise, the least popular top preference is deleted from all ballots and the process repeats.
\end{itemize}
IRV is also not Pareto efficient as a good candidate might be deleted early, if they fail to win many first choice votes. It is however remarkably resistant to strategic voting~\cite{bartholdi1991single} while retaining some properties that approval voting lacks, such as selecting the majority winner if one exists. This makes IRV particularly appealing when the community is asked to choose between alternatives in a non-binding way. The result can be further ratified by a referendum.

In some cases, IRV (and any voting system using ordinal preferences) might force the voters to inadvertently submit misleading information. For example, IRV assumes that the first and second place candidate on every ballot are separated by an equal amount, whereas some voters might be indifferent while others strongly in favour of their first choice only. Approval voting sometimes gets around this issue by asking for even less information. Ordinal preferences can be easily elicited by an \emph{auction} which is undesirable for an election. A better alternative is to use an ordinal voting mechanism such as majority judgment~\cite{balinski2011majority} or combine approval voting with \emph{token locking}: voters who feel strongly about some candidate may lock their vote tokens for longer, indicating that this election is particularly important to them.

\subsection{Confidentiality}
One of the initial goals of Bitcoin, as well as arguably the first design consideration when implementing a voting system on which the governance system will be based, is the approach to \emph{privacy}. While its definition is fairly intuitive, we make a distinction between \emph{secrecy} and \emph{pseudonymity}. 
\begin{definition}[Type 1: Secrecy]\label{def:secrecy}
A blockchain governance system satisfies secrecy if whenever a decision-making process is held, an adversary cannot guess the input of any participant better than an adversarial algorithm whose only inputs are the overall tally and, if the adversary is a participant, the adversary's input.
\end{definition}

This definition follows from the early work of 
Benaloh, cf. \cite{DBLP:conf/focs/CohenF85} 
and has been formally modeled in numerous subsequent works, e.g., see 
the model of \citet{juels2010coercion}. This is the strongest of the two notions and typically what would be required of an offline voting system (e.g., traditional elections in most countries). Often, true secrecy is difficult to accomplish in a decentralised setting or might be undesirable. For example, many blockchain combine on-chain governance with \emph{off-chain} elements, such as discussions on forums. These discussions may be part of the formal governance model and could be combined with an off-chain poll, based on the on-chain distribution of voting power. In these cases there could be a benefit in using \emph{pseudonyms}, keeping the real life identity safe but tying their public discourse with their actual vote. This is particularly relevant when the distribution of voting power distribution. Even though not explicitly mentioned by name, the Bitcoin whitepaper provides an explanation about why \emph{pseudonymity} \cite{nakamoto2008bitcoin} might be a good enough alternative.
\begin{definition}[Type 2: Pseudonymity]\label{def:pseudonymity}
A blockchain governance system satisfies pseudonymity if no participant is required to reveal their real-life identity to participate in the decision-making processes.
\end{definition}
The reason for the development of this notion is that blockchain systems are usually designed with the assumption that consensus is achieved 
\emph{only} with regards to the shared ledger; it is impossible to keep track of any information outside of it. Therefore, the same techniques used to keep track of the distribution of wealth (e.g., publicly announcing and linking transactions together), can be used to provide voting rights to the people actually involved in the blockchain without requiring much additional work. This is further related to the notion of \emph{suffrage}, which is defined in \Cref{subsec:suffrage}. For example, in Proof-of-Stake based cryptocurrencies like Cardano, voting rights for some applications are distributed based on the amount of \emph{stake} held by each user, as outlined in the paper by \citet{zhang2019treasury} describing the voting system used by the treasury system of that platform.  In practical terms, as long as the cryptographic information required when first producing one's online identity cannot be traced back to any real-life information, pseudonymity is satisfied. Privacy can be  further strengthened, considering the notion of \emph{coercion-resistance} \cite{juels2010coercion, cuvelier2013election}.
\begin{definition}\label{def:coercion-resistance}
A blockchain governance system satisfies coercion-resistance if whenever a decision-making process is held, a participant can deceive the adversary into thinking that they have behaved as instructed, when the participant has in fact made an input according to their own intentions.
\end{definition}
In a strict sense, this definition is arguably stronger than the guarantee provided by traditional elections: the voter should be able to deceive the adversary even about his participation, not just his vote. By definition, this exceeds the notion of privacy and requires at least one \emph{anonymous} channel of communication. Such a scheme is described in \cite{juels2010coercion}, but tallying requires an amount of communication which is quadratic in the number of votes. As such, this property is typically too demanding to be fulfilled in a blockchain setting, for most applications. However, it can be partially satisfied (e.g., if a ballot is encrypted in a way such that the voter can verify its inclusion when it is cast, but it is impossible for him to reclaim it later, if asked to prove that they voted in some way --- the fact that this only provides partial fulfillment of the property stems from the fact that if the participant's device leaks the random coins, then the ciphertext can be demonstrated to encode the participant's input).

\subsection{Verifiability}
To complement confidentiality, we now need a property that goes in the opposite direction,  namely \emph{verifiability}. This is a crucial property of every voting system, as it legitimises the election result. The widely accepted ``golden standard'' of verifiability is expressed below in the form of end-to-end verifiability. 

\begin{definition}[End-to-End Verifiability]
A blockchain governance system is verifiable if whenever a decision-making process takes place, participants are to able to verify their inputs were properly tallied and independent observers are able to verify that inputs from eligible participants were properly tallied.
\end{definition}
Furthermore, \citet{gharadaghy2010verifiability} split the definition of verifiability into two separate notions.
\begin{itemize}
    \item \textbf{Individual Verifiability:} It is possible for the voter to audit that his/her vote has been properly created (in general encrypted), stored, and tallied.
    \item \textbf{Universal Verifiability:} Everyone can audit the fact that only votes from eligible voters are stored in a ballot box, and that all stored votes are properly tallied.
\end{itemize}
At a high level, a system satisfying both properties would be called end-to-end verifiable -- but we refer to \cite{DBLP:conf/sp/CortierGKMT16} for more details on the notion of verifiability as well as the subtleties that arise in defining the concept formally. 

Intuitively, satisfying privacy (and \Cref{def:secrecy} in particular) as well as coercion-resistance \cref{def:coercion-resistance} should make verifiability more difficult to achieve. After all, these two limit the amount of information that a third-party could elicit by observing the blockchain. Despite this, it is indeed possible to achieve both to a certain adequate level. As exemplary schemes we can point to the work of  \cite{juels2010coercion} mentioned earlier, but also schemes such as the early work of Benaloh and Tuinstra \cite{DBLP:conf/stoc/BenalohT94}, 
the Benaloh-challenge approach  \cite{DBLP:conf/uss/Benaloh06} that has influenced  a lot of practical e-voting systems, see e.g.,  \cite{DBLP:conf/pkc/Kiayias0Z17}, or the hardware token based approach of \cite{DBLP:conf/crypto/AlwenOZZ15}. This latter work also provides a  comprehensive modeling of the concept of incoercibility that extends well beyond the setting of e-voting per se and can be immediately applicable to the blockchain setting as well. 

\subsection{Accountability}\label{sec:accountability}
The quest for accountability in governance is not a recent pursuit, as it was clearly recognised by the ancient Egyptians and the ancient Greeks \cite{dykstra_1939}. Since then, accountability as a concept has been split into multiple types and dimensions. For example,   \citet{grant2005accountability} outlines that accountability can take two general forms: vertical (where a party is accountable to other parties that are higher in a given hierarchy) and horizontal (where a party is accountable to other parties that are not higher or lower in a given hierarchy). Although \textit{collective} accountability is often implicitly implied in coin-based voting, \textit{individual} accountability is not. That is, if enough voters vote for a bad decision, the coin value of every voter declines whether or not they supported the decision. Individual accountability can take various forms, the most prominent of which is often referred to as `skin in the game', where participants have an individual investment that will be directly affected by their individual actions.

Even though only the decision-makers take part in governance, accountability should capture the possible harm incurred to the community-members as well. This is an added layer of security required to align the incentives of these two types of participants, particularly in governance designs where the two groups could be disjoint (e.g., voting rights based on a governance token that has no other function or direct relation to any on-chain activity).
\begin{definition}
A blockchain governance system satisfies the  property of accountability if whenever participants bring in a change, they are held individually responsible for it in a clearly defined way by the platform.
\end{definition}
 Examples outside the blockchain space include the work done in \citet{skin_in_the_game}, where participants review publications and those having more `skin in the game' (evaluating publications in which they will be marked as co-authors) have an increased individual interest in ensuring that a study’s ambiguously reported methods and analyses are clarified prior to submission. Examples in the blockchain space include Polkadot's governance system \cite{polkadot_governance}, where voters who vote in favour of a proposal will have their stake locked until the proposal is `enacted' or deployed.

\subsection{Sustainability}\label{sec:sustainability}
Changes in blockchain governance rely on two main actors: those who develop and propose the changes, and those who decide on whether or not to adopt these changes. Contributions from both actors help the platform to adapt and evolve and need to be rewarded.

\begin{definition}[Sustainable Development]
A blockchain governance system sustains development if it incentivises, via monetary rewards or otherwise, participants who develop successful improvement proposals for the platform.
\end{definition}

\begin{definition}[Sustainable Participation]
A blockchain governance system sustains participation if it incentivises, via monetary rewards or otherwise, participants who participate in the decision-making process of the platform.
\end{definition}
\begin{remark*}
Sustainability is different from accountability in both moral and practical terms.
Contrary to the definition of Accountability, Sustainability rewards development or participation with no regard to its outcome (ideally, before the respective agents have to perform the work or incur any costs). Accountability relates to possible penalties applied afterwards, once the effects of a particular change are apparent. For example, rewarding users just for voting would somewhat enable sustainable participation, but would not qualify for accountability. On the contrary, penalizing voters who approved a malicious proposal, without ever rewarding anyone, would only meet the definition of accountability.
\end{remark*}
The idea  behind having participation and development incentives in place is to \textit{help} justify the cost of engagement, which can lead to higher voter participation or more contributions to the platform. These incentives can take various forms, from monetary incentives to reputation- or merit-based incentives \cite{reputation_as_incentive}. However, Sustainable Participation could be a double edged sword if applied carelessly (e.g., \cite{panagopoulos2013extrinsic, shineman2018if}. A monetary reward that is too small might convert a moral decision into a financial one, paradoxically decreasing participation. While in general increased participation also leads to an increase in information acquisition from the voters, it is certainly more beneficial to have a smaller set of participants that have done their due diligence and vote as honestly as possible, than a larger group of disinterested individuals who cast votes at random just to collect rewards.

\subsection{Liveness}
In formal, on-chain governed platforms, the process for proposing and adopting changes is often constrained by fixed-length time periods. An example of this is Tezos's Granada protocol \cite{tezos_governance}, where a proposal has to go through five governance cycles (each lasting roughly two weeks) in order to be adopted. In such platforms, an unforeseen event that requires urgent action will not be resolved promptly through the platform's governance process. Therefore, a blockchain governance system must not only be able to process regular changes, but also urgent ones.
\begin{definition}
A blockchain governance system satisfies liveness if it is capable of incorporating an input of urgency from the stakeholders and then being capable of acting on it in the sense that if an issue is deemed to be urgent according to some function, then the decision making procedure is capable of terminating within a reasonable amount of time, as a function of the urgency of the matter.
\end{definition}
This definition includes having some protection against \emph{denial of service} attacks, that would prohibit governance mechanisms from terminating in time. All systems evaluated in this work are safe, at least from a high level standpoint, ignoring implementation details.

Events like the DAO hack \cite{dao_hack_report} have shown the need for blockchain governance systems to be able to accommodate inputs of urgency and act on them within a suitable amount of time. An example of blockchain governance system with liveness measures is Polkadot \cite{polkadot_governance}, which allows for emergency referenda to be initiated by an assigned technical committee. Others, such as MakerDAO, implement an emergency shutdown functionality: since it is running on Ethereum, in an emergency the smart contact can suspend its normal operation and return the invested assets to their owners.

\section{Evaluations}\label{section:evaluations}
In this section, we evaluate a number of popular platforms with respect to the properties outlined in Section \ref{section:properties}. The platforms below were chosen such that they present an overview of current approaches. An overall view of the evaluations can be found in Table \ref{tab:evaluations}. We start with Bitcoin and Ethereum, two of the oldest and most influential blockchains. These two use proof-of-work for consensus and rely mostly on their developers for governance, who maintain a connection with the community but ultimately have control over the direction of the platform. Continuing, we consider Tezos, Polkadot and Decred. The first two use proof-of-stake, while Decred takes a hybrid approach. In particular, whereas Tezos and Decred favour ``direct'' democracy, Polkadot uses a \emph{council} as well, representing two fundamentally different approaches to managing how voters express their preferences and interact with the governance process. Next, we study Project Catalyst and Dash, which incorporate a treasury in their decision making, meaning that the result of the voting process needs to respect a budget. Finally we consider Compound, Uniswap and MakerDAO that use a governance token approach. In the case of Compound and Uniswap this token is purely used for voting, while for MakerDAO it also supports the normal operation of the Maker protocol.

Gathering all the necessary information about every governance system is not always easy: typically, the platform's white paper would contain a very high level overview. Moore details can sometimes be found on the websites of the respective blockchains, but often the complete picture can only be acquired by interacting with a wallet, voting app or forum. Keeping that in mind, we have made our best efforts to cite the relevant sources. 

\begin{remark*}
In the main text we only include a high-level description and evaluation of the governance protocols. A more in-depth study, along with a point-to-point comparison with respect to each property can be found in the appendix.\\

\noindent
The information provided is accurate as of April, 2022.
\end{remark*}

\begin{table*}
  \centering
  \begin{tabular}{p{2.0cm}p{0.1cm}p{0.1cm}p{0.1cm}p{0.1cm}cp{0.1cm}p{0.1cm}ccp{0.1cm}p{0.1cm}c}
  \hline
  \textbf{Platform} 
  &\multicolumn{4}{p{2cm}}{\textit{Suffrage}}
  &\textit{Pareto Efficiency}
  &\multicolumn{2}{p{1.5cm}}{\textit{Confidentiality}}
  &\textit{\quad Verifiability}
  &\textit{Accountability} 
  &\multicolumn{2}{p{1.5cm}}{\textit{Sustainability}} 
  &\textit{Liveness} \\ \hline
    \\ 
    & \rotatebox{45}{\textit{Identity-based}} & \rotatebox{45}{\textit{Token-based}} & \rotatebox{45}{\textit{Mining-based}} & \rotatebox{45}{\textit{Meritocratic}} & & \rotatebox{45}{\textit{Secrecy}} & 
    \rotatebox{45}{\textit{Coercion Resistance}} & & &
    \rotatebox{45}{\textit{Development}} &  \rotatebox{45}{\textit{Participation}} & \\ 
    
    \textbf{Bitcoin} & \pie{0} & \pie{0} & \pie{360} & \pie{180} & \pie{0} & \pie{90} & \pie{0} & \pie{270} & \pie{0} & \pie{0} & \pie{0} & \pie{180}\\
    \textbf{Ethereum} & \pie{0} & \pie{0} & \pie{0} & \pie{180} & \pie{0} & \pie{90} & \pie{0} & \pie{270} & \pie{0} & \pie{0} & \pie{0} & \pie{180}\\
    \textbf{Catalyst} & \pie{0} & \pie{360} & N & \pie{90} & \pie{180} & \pie{270} & \pie{90} & \pie{270} & \pie{90} & \pie{360} & \pie{360} & \pie{0}\\
    \textbf{Dash}  & \pie{0} & \pie{360} & \pie{0} & \pie{0} & \pie{180} & \pie{90} & \pie{0}  & \pie{360} & \pie{0} & \pie{360} & \pie{360} & \pie{0}\\
    \textbf{Tezos} & \pie{0} & \pie{360} & N & \pie{0} & \pie{180} & \pie{90} & \pie{0}  & \pie{360}  & \pie{0} & \pie{0} & \pie{0} & \pie{0}\\
    \textbf{Polkadot} & \pie{0} & \pie{360} & N & \pie{90} & \pie{360} & \pie{90} & \pie{0}  & \pie{360}  & \pie{360} & \pie{180} & \pie{0} & \pie{360}\\
    \textbf{Decred} & \pie{0} & \pie{360} & \pie{0} & \pie{0} & \pie{180} & \pie{90} & \pie{0}  & \pie{360} & \pie{0} & \pie{360} & \pie{0} & \pie{0}\\
    \textbf{Compound} & \pie{0} & \pie{360} & \pie{0} & \pie{90} & \pie{180} & \pie{90} & \pie{0} & \pie{360}  & \pie{0} & \pie{0} & \pie{0} & \pie{360}\\
    \textbf{Uniswap} & \pie{0} & \pie{360} & \pie{0} & \pie{90} & \pie{180} & \pie{90} & \pie{0}  & \pie{360} & \pie{0} & \pie{0} & \pie{0} & \pie{180}\\
    \textbf{Maker DAO} & \pie{0} & \pie{360} & \pie{0} & \pie{0} & \pie{270} & \pie{90} & \pie{0}  & \pie{360} & \pie{0} & \pie{0} & \pie{0} & \pie{360}\\
  \end{tabular}
  \caption{Overview of the evaluations of each property against 
  each of the chosen platforms.\\ \normalfont{Every platfom might satisfy each 
  property to a different degree. This is shown by a filled circle for robustly 
  meeting the definition down to an empty circle if clear improvements are 
  needed. The letter $N$ is used if a property does not apply.}}
  \label{tab:evaluations}
\end{table*}

\subsection{Bitcoin}
Bitcoin \cite{nakamoto2008bitcoin} is the most prominent blockchain platform and it is a proof-of-work, mostly off-chain governed blockchain. The Bitcoin Improvement Proposal (BIP) process \cite{bip0002} is Bitcoin's primary mechanism for `proposing new features, for collecting community input on an issue, and for documenting design decisions'. An individual or a group who wishes to submit a BIP is responsible for collecting community feedback on both the initial idea and the BIP before submitting it to the Bitcoin mailing list for review. Following discussions, the proposal is submitted to the BIP repository as a pull request, where a BIP editor will appropriately label it. BIP editors fulfil administrative and editorial responsibilities. There are repository `maintainers' who are responsible for merging pull requests, as well as a `lead maintainer' who is responsible for the release cycle as well as overall merging, moderation and appointment of maintainers \cite{bip0002_contributing}. Maintainers and editors are often contributors who earnt the community's trust over time. A peer review process takes place, which is expressed by comments in the pull request. Whether a pull request is merged into Bitcoin Core rests with the project merge maintainers and ultimately the project lead. Maintainers will take into consideration if a patch is in line with the general principles of the project; meets the minimum standards for inclusion; and will judge the general consensus of contributors \cite{bip0002_contributing}. 

There are stages through which a BIP can progress, including `Rejected' and `Final'. In progressing to a status of `Final', there are two paths:
\begin{itemize}
    \item \textit{Soft-fork BIP}. A soft-fork upgrade often requires a $95\%$ miner super-majority. This is done via an on-chain signalling mechanism introduced in \cite{bip0009}. 
    \item \textit{Hard-fork BIP}. A hard-fork upgrade requires adoption from the entire `Bitcoin economy', which has to be expressed by the usage of the upgraded software.
\end{itemize}

\noindent {\textbf{Evaluation.}}
It is important to note here that the Bitcoin decision-making mechanism is informal, at least with respect to other platforms. Clearly, the on-chain aspects of Bitcoin's governance satisfy pseudonymity, but not secrecy or coercion resistance as no `votes' are even encrypted. The same is true for its off-chain component. This has the advantage that the system is mostly verifiable, even though having part of the deliberation take place in public forums is harder to track and could be an impermanent storage solution. Since the decision-making process is informal, without clearly defined structure or voting rules, Pareto Efficiency (to any degree) cannot be guaranteed. Sustainability and Accountability fail for the same reason, as there are no defined rules for either. Liveness is arguably partially satisfied, given the informality and flexibility of the BIP system. Since miners are guaranteed to explicitly signal their approval or disapproval of soft-fork upgrades \cite{bip0009}, mining-based suffrage is satisfied. Although those with previous positive contributions and relevant expertise are able to provide substantial inputs in the decision-making process, there is no explicit guarantee of their decision-making rights due to the informality of the process. Despite this, we conclude that meritocratic suffrage is \textit{likely} satisfied.

\subsection{\textbf{Ethereum}}
\label{sec:ethereum}
Ethereum \cite{eth_white_paper} is one of the most significant second-generation blockchain platforms. It is proof-of-work and governed off-chain, using the Ethereum Improvement Proposal (EIP) process \cite{eip0001} as a mechanism for proposing and integration changes. It is almost identical to that of Bitcoin, without giving miners the option to signal their preferences on-chain.

\subsection{Tezos}
Tezos \cite{tezos_docs} is a proof-of-stake, on-chain governed blockchain platform, which defines its governance process as `self-ammending''. Contrary to Bitcoin or Ethereum, participating in governance is based on \emph{stake}. Specifically, Bakers (also known as \emph{delegates}) need to have at least $8,000$ XTZ (called a \emph{roll}) and the infrastructure to run a Tezos node in order to gain \emph{both} block producing and voting priviledges. Community members who have fewer than $8,000$ XTZ or are unwilling to spend the computational resources can \emph{delegate} their stake to bakers, who produce blocks and vote on their behalf. The voting process is currently divided in five governance periods, each period spanning roughly two weeks: Proposal, Testing-vote, Testing, Promotion-vote and Adoption. During the proposal period, \emph{approval voting} is used to select the winning proposal, which must also be accepted by at least $5\%$ of the total vote. In testing-vote and promotion-vote the possible options are `Yea', `Nay' or `Pass'. A quorum between $0.2$ and $0.7$ of the total stake need to be reached, and the proposal is implemented if an $80\%$ supermajority of `Yea' is reached.

{\textbf{Evaluation.}}
As with Bitcoin, Tezos only satisfies Pseydonymity, but is completely verifiable. Pareto Efficiency is more nuanced. If a proposal receives less than $5\%$ of the upvotes or is tied with another proposal, no proposal will pass, even though operators could have voted for some proposals. However, given the properties of approval voting outlined in \Cref{sec:Pareto}, this effect is mild. In addition, the selected outcome is checked once again at the last step. Pareto efficiency could be further hampered under the assumption that the proposals appearing in a single voting period are \emph{too many} or \emph{too technical} to evaluate in the allotted time, before the vote. This could make voters inadvertently split their votes and abstain on many proposals, either leading to a deadlock if no proposal reaches $5\%$ or favoring \emph{whales} (i.e. users with many tokens). To see this, consider that between 3 proposals $A, B$ and $C$ one whale with $40\%$ of the tokens favours $A$ while every other user equally likes $B$ and $C$, but dislikes $A$. If the whale votes in favour of $A$ and the other voters evenly split their votes between $B$ and $C$, $A$ could win the election. A possible solution to this would be to separate \emph{vote} from \emph{stake} delegation. Voters could transfer their voting rights to more knowledgeable individuals that they trust which could consolidate their votes, while retaining their block production capabilities. Accountability or Sustainability are not satisfied. Given the lack of flexibility of the on-chain governance model, the Tezos governance system is incapable of taking inputs of urgency. Although a Gitlab issue or a pull request could be initiated without going through the formal on-chain route, it is still not the officially documented, and certainly not the `self-amending'', way by which the system processes inputs.

\subsection{Polkadot}
Polkadot \cite{polkadot_governance} is a proof-of-stake,  {\em mostly-on-chain} governed block\-chain platform with a number interesting additions, including an elected council and a technical council. Voters require at least $5$ DOT to participate in governance and their voting power is based on stake. At a glance, the voters elect councillors, directly vote on referenda and submit proposals. The councilors then have the power to \emph{veto} dangerous proposals, elect the technical committee, submit proposal of their own for approval by the voters and also control the \emph{treasury}. The technical council can submit \emph{emergency} referenda, that are implemented immediately if approved.

More specifically, the council consists of 13 members with 7 day tenures. They are elected using an approval voting based method, the weighted Phragmén election algorithm (e.g. \cite{brill2021phragmens}. An in-house refinement of Phragmén called Phragmms \cite{cevallos2021verifiably} could be used in the future. During a referendum election, an \emph{adaptive} quorum is used, requiring a different majority and turnout based on  how the referendum was created (e.g, by the community or a weak council majority). A successful referendum enters a 28 day waiting period before enactment, unless it is an emergency. Typically, the votes cast are \emph{locked} for these 28 days. However, the voters can increase their voting power by voluntarily locking them for longer (or decrease it by not locking at all). The treasury is controlled by the council, which decides whether to allocate funds to proposals that ask for them based on current supply.

{\textbf{Evaluation.}}
As usual, only pseudonymity and verifiability are satisfied. Council elections and referenda voting functions are Pareto efficient. In addition, the voters have the ability to lock their votes for an extended time, to signal the strength of their preferences. Arguably, a veto might not be Pareto efficient if there is $100\%$ consensus in a referendum. However, this is an extremely contrived case. Voting in favour of a proposal requires funds to be locked in until the proposal is enacted. The documented rationale behind this is to hold voters responsible for a proposal that they vote for, satisfying accountability and further reinforcing Pareto Efficiency. There are no explicit or direct rewards given for participation or contribution to satisfy sustainability. However, Polkadot have deliberately chosen \emph{against} monetary rewards for voters, for justified \href{https://polkadot.network/blog/a-walkthrough-of-polkadots-governance/}{reasons}. Often the rewards for voters are too low for a significant effect, as detailed in \Cref{sec:sustainability}. However, council members should probably receive some direct compensation. Even though their tenure is short, they hold a lot of power and should have the ability to devote themselves full time. The Polkadot governance mechanism is capable of taking in inputs of urgency (i.e. emergency referenda) and acting on it if deemed urgent by the council, all whilst being able to terminate within an amount of time proportional to the urgency. Token-based suffrage \emph{is} satisfied since only token holders are allowed to vote. The council adds teams to the technical committee (which is able to propose emergency referenda) based on their positive technical contributions and expertise. However, those teams are chosen by council members only and a positive contribution does not equate to a guarantee of an input in a decision-making process.

\subsection{Decred}
Decred is a hybrid proof-of-work and proof-of-stake system that is mostly on-chain governed \cite{decred_documentation}. Voters can participate in governance by locking enough DCR, which is the native token of Decred. This provides them with \emph{tickets} which supplement the consensus protocol and can also be used for voting. High level issues that require funds from the Decred Treasury are handled off-chain, in \href{https://proposals.decred.org/}{Politeia}. This deliberation results in an election which is cryptographically coupled to a snapshot of the chain. A $20\%$ quorum is needed, with over $60\%$ of the votes being in favour. The on-chain component is the Decred Change Proposal (DCP) \cite{decred_change_proposals}, through which the consensus mechanism is updated. This requires a $10\%$ quorum and $75\%$ majority of approval. Failing to meet the quorum, the election will be repeated in the next cycle. If it is successful, a `lock-in' period begins, after which all nodes should update their software.

\textbf{Evaluation}
The votes are not encrypted, therefore only pseudon\-ymity and verifiability are satisfied. Pareto efficiency is somewhat satisfied: there are similar issues as Tezos, but the added role of Politeia could improve the outcome. Sustainable development is satisfied (somewhat informally) but there are no specific rewards for participating in governance. Voters receive rewards, but these have to do with their role in the hybrid consensus protocol. Accountability could be improved, as the token locking required for voting is shorter than the timelock for successful proposals.

\subsection{Compound}
Compound~\cite{compound_white_paper} is a protocol running \emph{on} the Ethereum blockchain that establishes money markets. Governance in Compound is fuelled by an ERC-20 compatible token called \href{https://etherscan.io/token/0xc00e94cb662c3520282e6f5717214004a7f26888}{COMP} \cite{comp_coinbase}. These \emph{governance} tokens are distributed to the community through various channels: some are allocated to users based on their invested assets, others to Compound Labs Inc. shareholders and employees, etc. Holding COMP allows users to vote, delegate to others and create proposals, which are executable pieces of code. Once submitted, these proposals enter a two day review period, following a three day election. A proposal is successful if a majority is in favour and a quorum is reached. After that, the proposal is \emph{locked} for two days before implementation, for security reasons. In addition, the \emph{Pause Guardian} (controlled by a community appointed multi-signature) can suspend most functionalities of Compound at any time.

\noindent{\textbf{Evaluation}}
Every step of the governance process is performed by interacting with smart contracts on Ethereum, without any further cryptographic techniques, satisfying pseudonymity and verifiability. Once a proposal enters the voting phase, the voters only have two options: yes or no, which is clearly Pareto Efficient. If there are multiple incompatible options (e.g., values of a specific parameter), these proposals would have to be dealt with sequentially: the actual order could bias voters, which complicates their decisions and leaks information. Therefore, Pareto Efficiency is somewhat satisfied (e.g., between two highly popular proposal, the slightly less popular one might win if it is up for election first and then the users might be less eager to implement another change). Once a proposal is executed, its creator and voters are completely independent from its future and there are no rewards associated with the process. Therefore, neither availability or sustainability are satisfied. The total time between creating a government proposal and voting for it takes 7 days, 2 of which are hard-coded into the Timelock. This window for immediate action is only open right after a vote, but adding the Pause Guardian, liveness is satisfied.
Since voting eligibility depends only on having COMP tokens, which can be exchanged and are initially distributed to addresses with assets on Compound, token-based suffrage is satisfied. Some COMP tokens are distributed or reserved for members of the Compound team. Therefore, meritocratic suffrage is slightly satisfied.

\subsection{\textbf{Uniswap}}\label{sec:uniswap}
We briefly sketch Uniswap~\cite{uniswap_white_paper} governance, which combines off and on-chain components. The on-chain part of its governance system is almost identical to Compound~\cref{sec:compound}, using the UNI token instead. However, UNI can also be used to empower off-chain processes. The off-chain discourse takes place on the Uniswap governance \href{https://gov.uniswap.org/}{forum}, where 2 types of posts have particular significance. The first is the Temperature Check, whose goal is to gauge interest in changing the status quo. After 3 days there is a poll, where users have vote according to the amount of UNI they hold on-chain. If a majority is reached and quorum are reached, a Consensus Check is created on the same forum. During the 5 day duration of the Consensus Check, a proposal needs to be fleshed out. In the end, a second poll is brought before the users, this time possibly containing many alternatives. As long as the highest ranked alternative receives more than 50,000 UNI, an on-chain Governance Proposal is created and handled like in Compound.

\subsection{Maker DAO}
Maker DAO~\cite{makerdao_white_paper} is a decentralized organization running on Eth\-ereum and based on the Maker Protocol. One of its novel features is using a two-token system, with DAI, which is a stablecoin pegged to the U.S. dollar, and MKR which the governance token. MKR serves an additional purpose however: to support DAI's peg. The governance system employs both on and off-chain elements. The off-chain component takes place at the Maker DAO \href{https://forum.makerdao.com/}{forum}, where users can create Forum Signal Threads, which are followed by a poll. Each forum user has a single vote, irrespective of MKR. These are further ratified on-chain by Governance Polls, which employ \emph{instant-runoff voting}, weighted by the MKR of each voter. Finally, changes to the protocol (which are pieces of executable code) are enacted by Executive Votes. These follow a \emph{continuous} approval vote system, with the most approved Vote at any given time being the actual implementation. For security reasons, these changes happen after a 24 hour waiting period and there is also an emergency shutdown functionality, triggered if the community locks enough MKR.

\noindent{\textbf{Evaluation.}}
As there is no vote encryption, only pseudonymity and verifiability are satisfied. Pareto Efficiency is improved compared to other designs by using instant-runoff voting to handle competing proposals, thus giving voter a richer action space to declare their preferences accurately (without requiring multiple rounds or additional strategic behaviour). Suffrage is also improved, as there is a clear connection between MKR tokens and the overall functionality of Maker DAO, further coupling its value to some actual generated utility.

\begin{remark*}
Project Catalyst and Dash also include a \emph{treasury}, which complicates the voting process. Funds are periodically collected by the normal blockchain operation and allocated to fund its development and undertake projects \emph{whose results may take months to materialize}. In addition, at every funding round more than one proposal may be selected, as long as their total cost does not exceed a budget. Voters need additional flexibility to signal their preferences. Specifically, they need to compare a proposals perceived value with it with its budget and think about the opportunity cost of funding it. This is closely related to the field of \emph{Participatory Budgeting} (e.g., \cite{participatory_budgeting, benade2021preference, buterin2019flexible}). Decred also includes a treasury. The salient difference (e.g., with Project Catalyst) is that competing proposals are first debated off-chain, rather than set to compete on-chain for some portion the budget available in one round of funding. The final vote \emph{is} on-chain, but only as a referendum on proposals that already acquired off-chain support.
\end{remark*}

\subsection{\textbf{Project Catalyst}}
Project Catalyst \cite{catalyst_community} is the on-chain treasury governance system used by the Cardano blockchain, which is proof-of-stake. Governance takes place in twelve week periods called funds and involves a number of additional agents, on top of the usual voters, whose voting power and eligibility is dependent on stake ownership. At the beginning of the fund, community generated proposals (which include a corresponding budget) are submitted. These are then reviewed by Community Advisors (CA's) and these reviews are further checked for their quality by veteran Community Advisors (vCA's), both of which are rewarded for their efforts. Given these evaluations, an approval voting based mechanism~\cite{zhang2019treasury} is used. The proposal whose `Yes' votes minus the `No' votes are more than $5\%$ of the total votes received is eligible for funding. These eligible proposals are then sorted according to their approval. If the available funds are not enough to cover some proposal, it is skipped and a less popular (but cheaper one) could take its place. In addition, there is the Catalyst Circle~\cite{catalyst_circle}, an elected group of representatives that oversees Catalyst and a delegated voting system is proposed for future iterations.

{\textbf{Evaluation.}}
Everyone participates in Project Catalyst using their wallet address. Voters submit \emph{encrypted} ballots (padded with some randomness), using the public key issued by a committee, which tallies the votes and decrypts the result. If the voter address is linked to a real identity, the only information available is that this particular person voted, keeping the contents secret. The ballot itself cannot be decrypted by the voter and if the random padding is not kept, it is impossible even for the voter to convince anyone of the way they voted. The result of the vote can be independently verified and long as the voter saved the random padding, they can verify that their particular vote was counted. Therefore, there is a (somewhat contrived) sequence of events after which a voter would be unable to check that their ballot has been added.

In some cases, proposals with fewer votes will be prioritised for their lower budgets. For example, if the total fund is 100 and the three winning proposals have budget 1, 50 and 50 (in order of popularity) the last proposal will not receive funding, even though every voter might prefer funding the two 50 proposals. Additionally, each voter could submit an uninformative `no' vote to many proposals, in order to maximize the winning chance of their favourite. A potential mitigation would be to use techniques from Participatory Budgeting~\cite{benade2021preference} and Distortion~\cite{anshelevich2021distortion}, which use a small amount of \emph{ordinal information} (e.g., asking voters to compare between 2 proposals or to list their most favourite one) to improve the quality of the outcome. Overall, Pareto Efficiency is only \emph{somewhat} satisfied.
    
There are no explicit, on or off-chain, penalties. Proposers need to submit progress reports about their projects to keep receiving funding and community advisors can be penalized for poor reviews or absence. As these are either centralized or community-driven without clearly described mechanisms, accountability is mostly \emph{not} satisfied. Although there is no explicit reward given to the proposer, it is her responsibility to request the amount which cover the cost of her work. All other parties are rewarded for participating in the governance process and to an extent receive larger rewards for additional effort. Each Project Catalyst Fund follows a 12 week timeline. Liveness is not satisfied: even though the funds can be released in accordance with each proposal's progress, there is no mechanism to take urgent action. Voting rights depend only on having at least $500$ ADA. There are no guaranteed voting rights based on previous positive contributions, however, community advisors can affect the outcome of the votes through their reviews.

\subsection{\textbf{Dash}}
Dash \cite{dash_docs} uses proof-of-work for the underlying consensus mechanism, but includes an additional layer of functionality enabled by \emph{masternodes}, including governance and treasury fund allocation. These masternodes are users that have locked at least $1,000$ DASH (called collateral, which is part of their stake) and also operate a server. The treasury operates in similar fashion to Project Catalyst, but requires a $10\%$ difference between `Yes' and `No' votes for eligible proposals. Proposals can be submitted by anyone, but require spending $5$ DASH to ensure that only serious enough issues are raised. Only masternodes may vote and there are no designated roles for reviewers or elected representatives. Additionally masternode do not collect rewards specifically for voting, but are rewarded for the entirety of their duties.

{\textbf{Evaluation.}}
The system only satisfies pseudonymity and verifiability, as votes are public. Pareto Efficiency is similar to Project Catalyst. Although masternodes have collateral, this is not directly tied to governance and could be withdrawn immediately after enacting some controversial proposal. Sustainable development is satisfied through the treasury, but sustainable participation could be improved as the masternode rewards are not specific to voting, but conensus as well. Additionally, there is an issue of Suffrage since \emph{only} token holders (having at least 1,000 DASH, or about $54,000\$$) who are also willing to run a server can participate, leaving other token holders without representation.

\section{Challenges \& Research Directions} \label{sec:open} 

It should be clear from our exposition so far  that the blockchain governance space is still rife with challenges and open questions. 
We summarize in this section a number of them to motivate future research in the area. 

{\em I. Tradeoffs between Privacy vs. Verifiability and Suffrage.} 
The tension between verifiability and privacy stems from requirements
such as universal verifiability which mandates 
tracing  each decision back to the inputs of  decision-makers as determined by suffrage. The higher degree of privacy that is required, 
the more difficult it is to ensure verifiability; as a simple example from classical elections,
if the electoral roll remains private, then it is 
difficult for an external observer to verify whether the correct set of 
decision-makers has participated. This also creates a tension with suffrage
as types of suffrage that maximize inclusion, for the sake of verifiability, might have to expose a larger set of
community-members that otherwise would have remained private. Technically reconciling these properties is highly
non-trivial, especially if privacy aspects such as coercion resilience are desired. 

{\em II. Proofs of Personhood, Identity-based suffrage and tradeoffs with Privacy .}
While there is wide agreement that individual users should have  equal weight
in decision-making (something advocated in the context of election reform
for centuries, cf. \cite{howell1880}), achieving this type of suffrage
is particularly challenging in the context of decentralized systems. 
Even though some initial work is undertaken in this direction  e.g., \cite{proof_of_personhood}, and there are also connections with other
concepts in cyber-security such as CAPTCHAs \cite{DBLP:conf/eurocrypt/AhnBHL03}, nevertheless 
 the problem of achieving a satisfactory level of
 identity-based suffrage in the context of blockchain governance is still wide open.
 This challenge should be also considered from the lens of privacy,
 since in many cases of such proofs, community-members would have
 to reveal personally identifiable information to other actors 
 something that comes inevitably with privacy implications. 

{\em III. Meritocratic suffrage and tradeoffs with privacy. }
The challenge in the context of meritocratic suffrage is in two 
levels, first, in quantifying what type of merit itself should warrant
participation to decision-making. The second
level is recording reliably the relevant actions of community-members 
in the system so that it can be acted upon during the decision-making process. 
Finally, as in the case of proofs of personhood, there can be privacy
implications. Some early works in this direction show that 
privacy and merit may be reconciled, see e.g., the signatures of reputation primitive
\cite{DBLP:conf/fc/BethencourtSS10} but still, significantly more work is required to 
fully tackle the full spectrum of possible ways to express and act on merit. 

{\em IV. Exchanges, venture capital investors and token-based suffrage.} 
In the setting of token-based suffrage, an important consideration
is the fact that token-holders may choose custody solutions 
for their tokens for a variety of reasons (reducing risks regarding 
loss of keys, or the 
ability to access services or rewards provided by custody operators). 
While among some cryptocurrency users this is frowned upon (the tenet ``not your keys, not your coins'' is frequently repeated in social media) there is a large number of users that prefer to keep their digital assets in third party providers' systems.\footnote{Indicatively, statistics from the web-site https://cryptoquant.com/, at the time of  writing (May 2022),  suggest that about $13.3\%$ of the Bitcoin supply is held on exchanges. The figure for Ethereum is higher at slightly above $20\%$. }
This state of affairs, results in entities with inflated leverage in a token-based system that in some cases can control a very significant portion of the token supply. A related issue is the presence of venture capital firms that are early investors in some platforms and receive a large amount of tokens at preferential prices in exchange for funding initial development efforts. This similarly may result in increased leverage which can be perceived as unfair by other community-members.  

{\em V. Rational ignorance and inaction.} Rational ignorance \cite{rational_ignorance} is when decision-makers refrain from acquiring the knowledge required of meaningful input when voting, or when delegating their vote, due to the fact that the cost of acquiring that knowledge exceeds any expected potential benefits. A similar argument can be applied to developing improvement proposals, where inaction can be more rational than action if the cost of development (or even the act of preparing a proposal)
exceeds any potential benefits. 
These issues pertain to the property of sustainability which so far lacks a comprehensive theoretical framework in the context of blockchain governance. 
For some recent work that can be helpful in this direction see \cite{prato-wolton2016,prato-wolton2018}. 

{\em VI. Tradeoffs between accountability and utility.} 
Recall that  making decision-makers accountable suggests some degree of ``skin-in-the game'' on their side and  the natural way to achieve this suggests
some form of  restriction of the functionality that  is offered to them by the platform. 
As a result, the immediate utility that decision makers
can extract from the platform is reduced --- recall the  example of ``token lockup'' for the duration of a certain decision
making process. The main  challenge in this setting is to model
and quantify the relevant aspect of this utility reduction and mapping the spectrum of possible options so that the
the right balance between accountability and utility can be determined on a case by case basis.  

{\em VII. Tradeoffs between Liveness vs. Pareto Efficiency and Suffrage.}
As we discussed in the context of liveness, expedient decision-making is highly desirable. Unfortunately 
high expediency can come at odds with Pareto efficiency: if decision-makers have preferences which are not recorded
due to the system not giving them enough opportunity to them for  reacting, then it is easy to see that this can violate Pareto efficiency
(observe here that abstaining can be also a preference - however there is a distinction between having an actual preference and missing
the deadline to provide it to the system and preferring to abstain altogether).  Liveness can also exhibit a similar tradeoff with suffrage: 
the more exclusive the suffrage mapping from community-members to decision-makers is, the higher the expediency of the system 
may become - but this of course comes at the expense of the system being less inclusive. Striking the right balance between liveness
and these properties is another question on which future research should focus.

\section{Conclusion}

In this systematization work we focused on documenting  a comprehensive list of properties of blockchain governance. We took a first principles approach and derived seven fundamental properties using which we analyzed a number of widely used blockchain platforms. It is worth saying that 
there are also other platforms that we have attempted to cover, but these were either too poorly documented or were yet to implement governance mechanisms, thus we consider the list a comprehensive coverage of popular blockchain systems at the time of writing. 

The main outcome of the systematization effort, as illustrated in Table~\ref{tab:evaluations}, is that in many ways all current blockchain platforms either have deficiencies in their governance processes or allow significant room for improvement. It is worth also reiterating that achieving all stated properties to the highest possible degree  is impossible due to their conflicting nature and as a result it is inevitable that platforms must decide on appropriate tradeoffs between the various properties that are the most suitable for each particular setting. Arguably, without effective governance processes, blockchain technology will fail to reach its full potential. For one thing, software engineering practice has shown that software updates, extensions and patches are a necessity in the lifecycle of computer systems  and as a result, without proper governance, blockchain systems will fail to adapt to unanticipated use cases and mitigate software bug vulnerabilities that are inevitably discovered in any system. 

\section{Acknowledgements}
We would like to thank Yussef Soudan for his extensive research and participation in many meetings during the earlier stages of this work. Additionally, we thank Roman Oliynykov for providing many details and insights regarding the evaluation of Project Catalyst.

\bibliographystyle{IEEEtranN}
\bibliography{main.bib}

\appendix

\section{Further Details about Each Governance System}
Following our high-level overview in \Cref{section:evaluations}, we use the appendix to provide a more complete picture, including the finer details of each platform and how these affect each property.

\subsection{Bitcoin}
Bitcoin \cite{nakamoto2008bitcoin} is the most prominent blockchain platform and it is a proof-of-work, mostly off-chain governed blockchain. The Bitcoin Improvement Proposal (BIP) process \cite{bip0002} is Bitcoin's primary mechanism for `proposing new features, for collecting community input on an issue, and for documenting design decisions'. An individual or a group who wishes to submit a BIP is responsible for collecting community feedback on both the initial idea and the BIP before submitting it to the Bitcoin mailing list for review. Following discussions, the proposal is submitted to the BIP repository as a pull request, where a BIP editor will appropriately label it. BIP editors fulfil administrative and editorial responsibilities. There are repository `maintainers' who are responsible for merging pull requests, as well as a `lead maintainer' who is responsible for the release cycle as well as overall merging, moderation and appointment of maintainers \cite{bip0002_contributing}. Maintainers and editors are often contributors who earnt the community's trust over time. A peer review process takes place, which is expressed by comments in the pull request. Whether a pull request is merged into Bitcoin Core rests with the project merge maintainers and ultimately the project lead. Maintainers will take into consideration if a patch is in line with the general principles of the project; meets the minimum standards for inclusion; and will judge the general consensus of contributors \cite{bip0002_contributing}. 

There are stages through which a BIP can progress, including `Rejected' and `Final'. In progressing to a status of `Final', there are two paths:
\begin{itemize}
    \item \textit{Soft-fork BIP}. A soft-fork upgrade often requires a $95\%$ miner super-majority. This is done via an on-chain signalling mechanism introduced in \cite{bip0009}. 
    \item \textit{Hard-fork BIP}. A hard-fork upgrade requires adoption from the entire `Bitcoin economy', which has to be expressed by the usage of the upgraded software.
\end{itemize}
We now have an overview of the upgrades decision-making process in Bitcoin, which we will use to perform rough evaluations against the properties developed in Section \ref{section:properties}. It is important to note here that the Bitcoin decision-making mechanism is informal, at least with respect to other platforms. This results in rougher and less satisfying evaluations. 

\begin{itemize}
    \item \textbf{Suffrage}:
        Since miners are guaranteed to explicitly signal their approval or disapproval of soft-fork upgrades \cite{bip0009}, mining-based suffrage is satisfied. Although those with previous positive contributions and relevant expertise are able to provide substantial inputs in the decision-making process, there is no explicit guarantee of their decision-making rights due to the informality of the process. However, since meritocracy still does play a significant role in the process, we will conclude that meritocratic suffrage is \textit{likely} satisfied. 
        
    \item \textbf{Pareto Efficiency}. Since the decision-making process is informal, there is no defined voting rule, which specifies how the inputs result in a final outcome. Therefore Pareto efficiency is \emph{not} satisfied. 
    \item \textbf{Accountability}. The platform does not define any way by which it can hold participants responsible or accountable for their individual actions. Therefore, accountability is \textit{not} satisfied.

     \item \textbf{Confidentiality}: 
     \begin{itemize}
         \item \textbf{Secrecy}: Since the decision-making process among maintainers or reviewers is on public forums, an adversary might accurately guess each participant's input. Therefore, secrecy is \textit{not} satisfied. 
     
        \item \textbf{Pseudonymity}: There are no defined requirements for participants to reveal their identities. Some choose to participate with their real identities and others do not. Therefore, pseudonymity is satisfied.  

        \item \textbf{Coercion-resistance}: Since the deliberation process among maintainers and others takes place on public forums, an adversary might accurately guess each participant's input. Thus, coercion-resistance is \textit{not} satisfied.
    \end{itemize}
    \item \textbf{Verifiability}. The signaling mechanism used as a voting process for certain decisions is on-chain. However, even though the deliberation process takes place in public forums, the decision-making process remains informal, which makes it difficult to identify how inputs are incorporated from which parties and how they are tallied. However, such inputs can be traced through the public forums and any changes that are merged can be tracked on Github. Therefore, verifiability is \emph{mostly} satisfied.

    \item \textbf{Sustainability}:. There are no explicitly defined 
    incentives for contributors to develop BIPs.  Therefore, neither sustainable 
    \textbf{development} nor \textbf{participation} is satisfied. 
    	
    \item \textbf{Liveness}. Although no specific mention of inputs of urgency are provided by the platform, given the informality and flexibility of the BIP system, it is likely capable of taking inputs of urgency and acting on them in an amount of time that is a function of the urgency. Therefore, the platform \textit{likely} satisfies liveness.
\end{itemize}

\subsection{Tezos}
Tezos \cite{tezos_docs} is a more-recent proof-of-stake, on-chain governed block\-chain platform, which defines its governance process as `self-am\-mending''. In Tezos, to participate directly in the governance process, a participant is required to have at least $8,000$ tokens. A unit of $8,000$ tokens is called a \textit{roll} and it equates to a single vote. In this case, the participant is called a \textit{delegate}. Alternatively, to participate indirectly in the governance process, a participant can delegate whichever amount of tokens they have (which can be less than $8,000$) to an existing delegate. 

The voting process is currently divided in five governance periods, each period spanning roughly two weeks or 20480 blocks (i.e. 5 cycles). Note that for proposals to be submitted in Tezos, they need to be compiled without errors so that at the end of the governance process the proposal can be adopted automatically. The following is a breakdown of the five governance periods:
\begin{enumerate}
    \item \textbf{Proposal period}. Delegates can submit protocol amendment proposals using the proposals operation as long as the underlying codebase compiles with the change. Delegates then upvote their preferred proposal or proposals. The proposal with the most upvotes is selected. If there are no proposals, no proposals with upvotes of at least $5\%$ of the possible votes, or a tie between proposals, a new proposal period starts.
    
    \item \textbf{Testing-vote period}. Delegates can cast one vote to test or not the winning proposal using the ballot operation.
    
    \item \textbf{Testing period}. A test chain is forked for the entire testing period to ensure a correct migration of the context.
    
    \item \textbf{Promotion-vote period}. Delegates can cast one vote to promote or not the tested proposal using the ballot operation.
    
    \item \textbf{Adoption period}. The adoption period serves as a buffer time for users to update their infrastructure to the new protocol. At the end of this period, the proposal is activated as the new protocol and a new proposal period starts. Here, the Tezos node software is aware that at the end of this period it needs to update to the new protocol, hence why the governance process is described as `self-amending''. 
\end{enumerate}

In the \textbf{proposal period}, approval voting is used. In the \textbf{testing-vote} and \textbf{promotion-vote} periods, the voting method is as follows:

\begin{itemize}
    \item Each delegate can submit a single vote of a `Yea'', `Nay'' or `Pass''.
    \item If the participation reaches the current quorum and the proposal has a super-majority in favour, it goes through to the next stage.
    \begin{itemize}
        \item The quorum is the participation threshold, it has maximum value of 0.7 and a minimum value of 0.2, and it changes after every vote.
        \item A super-majority is when the number of `Yea'' votes is more than $80\%$ of the number of `Yea'' votes and `Nay'' votes summed together.
    \end{itemize}
\end{itemize}

Similar to the previously evaluated platforms, we perform the evaluations of the governance process in Tezos against the properties developed in Section \ref{section:properties}. 

\begin{itemize}
    \item \textbf{Suffrage}: 
        Only token-holders are able to vote, with or without delegation. Therefore, token-based suffrage \emph{is} satisfied.
        
    \item \textbf{Pareto Efficiency}.  If a proposal receives less than $5\%$ of the upvotes or is tied with another proposal, no proposal will pass, even though operators could have voted for some proposals. However, given the properties of approval voting outlined in \Cref{sec:Pareto}, this effect is mild. In addition, the selected outcome is checked once again at the last step. Therefore, Pareto efficiency is \textit{somewhat} satisfied. 
    
    \item \textbf{Confidentiality}: 
        \begin{itemize} 
        \item \textbf{Secrecy}: The delegation mechanism requires the public key of each delegate to be recorded on the ballot, and all ballots are public.  Therefore, secrecy is \textit{not} satisfied.
     
        \item \textbf{Pseudonymity}: Voters are not required to reveal their real-life identities to participate in the governance process; therefore pseudonymity is satisfied.

        \item \textbf{Coercion-resistance}. Since delegate votes (rolls) are tied to their chosen pseudo-identities, coercion-resistance is \textit{not} satisfied.
    \end{itemize}
    
    \item \textbf{Verifiability}. Since the votes and final tally are all public, verifiability is, by definition, satisfied.
    
    \item \textbf{Accountability}. Whether an operator is directly voting or delegating, the stake of each delegate is computed at the start of each voting period. This means that delegates can sell their stake before the adoption period ends and the proposal is activated. There are no accountability measures defined in Tezos. Therefore, accountability is \textit{not} satisfied.
    
    \item \textbf{Sustainability}: There are no explicit or direct incentives given 
    for 
    developing successful proposals or participating in the governance process. 
    Therefore, neither sustainable \textbf{development} nor 
    \textbf{participation} is satisfied. 
    \item \textbf{Liveness}. Given the lack of flexibility of the on-chain governance model, the Tezos governance system is incapable of taking inputs of urgency and responding to them in accordance to the severity of the issue. Although a Gitlab issue or a pull request could be initiated without going through the formal on-chain route, it is still not the officially documented, and certainly not the `self-amending'', way by which the system processes inputs. Therefore, liveness is \textit{not} satisfied.
\end{itemize}

\subsection{Polkadot}
Polkadot \cite{polkadot_governance} is a proof-of-stake, \textit{mostly-on-chain} governed block\-chain platform. To make any changes to the network, \textit{active} token holders and the \textit{council} administrate a network upgrade decision. Whether the proposal is proposed by the public (token holders) or the council, it  will go through a referendum to let all token-holders, weighted by stake, make the decision.

The council is an elected body of on-chain accounts that are intended to represent the passive stakeholders of Polkadot, currently consisting of $13$ members \cite{polkadot_governance}. The council has two major tasks in governance: (i) proposing referendums and (ii) vetoing dangerous or malicious referendums. The council implements what is called a \textit{prime member} whose vote acts as the default for other members that fail to vote before the timeout. The prime member is chosen based on a Borda count \cite{borda_count}. With the existence of a prime member, it forces councillors to be explicit in their votes or have their vote counted for whatever is voted on by the prime. The council also controls Polkadot's treasury and allocates funds to successful proposals.

Voting for councillors requires locking 5 DOT tokens (the native token of the platform)  and takes on an approval voting approach. A token-holder can approve up to 16 different councillors and the vote will be equalised among the chosen group, with each council term lasting 7 days. The approval voting method used is the weighted Phragmén election algorithm (e.g. \cite{brill2021phragmens}, where the candidates with most approvals are elected and, afterwards, a process is run that redistributes the vote amongst the elected set. This reduces the variance in the list of backing stake from the voters to the elected candidates in order to ensure that the minimum amount of tokens required to join the council is as high as possible. Running the Phragmén algorithm cannot be completed within the time limits of production of a single block. And waiting would jeopardise the constant block production time of the network. Therefore, as much computation as possible is moved to an off-chain worker, where validators can work on the problem without impacting block production time. An in-house refinement of Phragmén called Phragmms \cite{cevallos2021verifiably} could be used in the future.

A significant part of Polkadot's governance is the \textit{technical committee}, which is composed of teams that have successfully implemented or specified either a \textit{Polkadot runtime} or \textit{Polkadot Host} \cite{polkadot_governance}. These teams are added or removed from the technical committee via simple majority votes within the council. The technical committee can, along with the council, propose emergency referendums, which are fast-tracked for voting and implementation (e.g., for emergency bug fixes)

Besides electing councillors, token-holders get to vote in referendums. Each referendum has a specific proposal associated with it. Proposals can implement backward-compatible or backward-incompatible changes. Proposals can be submitted by token-holders, the council or the technical committee:
\begin{itemize}
    \item For token-holders to submit a proposal, a minimum amount of tokens must be deposited. If another token-holder agrees with the proposal, they can also deposit the same amount of tokens in the proposal's support. The proposal with the highest amount of bonded support will be selected to be a referendum in the next voting cycle. The referendum, in this case, will have positive turnout bias. That is, the smaller the amount of stake voting, the larger the super-majority necessary for it to pass \cite{polkadot_governance}. Specifically the proposal would pass if 
    $$
    \frac{\text{against}}{\sqrt{\text{turnout}}} < \frac{\text{approve}}{\sqrt{\text{electorate}}}.
    $$
    
    \item Proposals can only be submitted by the council through a majority or unanimously. In the case of a unanimous council, the referendum will have a negative turnout bias, that is, the smaller the amount of stake voting, the smaller the amount necessary for it to pass:
        $$
    \frac{\text{against}}{\sqrt{\text{electorate}}} < \frac{\text{approve}}{\sqrt{\text{turnout}}}.
    $$
    In the case of a majority, the referendum will be a majority-carries vote ($51\%$ of the votes is required to win). 
    
    \item The technical committee can propose emergency referendums subject to approval from the council. 
\end{itemize}

If a proposal passes in a referendum, then Polkadot’s logic automatically schedules it for enactment: autonomous enactment. This is unlike other systems where miners or validators often have unilateral power to prevent protocol changes by refusing to upgrade software. Proposals submitted by the council or token-holders are enacted $28$ days after the referendum, whereas ones submitted by the technical committee can be enacted immediately. 

To vote, a token-holder generally must lock their tokens up for at least the enactment delay period beyond the end of the referendum. This is in order to ensure that some minimal economic buy-in exists and to dissuade vote selling. It is possible to vote without locking at all, but the vote is worth a small fraction of a normal vote. It is also possible to voluntarily lock for more than one enactment period, in which case, the weight of the vote increases proportionally. This mechanism exists to ensure that users with little stake but strong opinions can express their conviction in referendums.

\begin{itemize}
    \item \textbf{Suffrage}: 
        Token-based suffrage \emph{is} satisfied since only token holders are allowed to vote. The council adds teams to the technical committee (which is able to propose emergency referenda) based on their positive technical contributions and expertise. However, those teams are chosen by council members only and a positive contribution does not equate to a guarantee of an input in a decision-making process. 
        Therefore, meritocratic suffrage is only \emph{slightly} satisfied.
    \item \textbf{Pareto Efficiency}. Council elections and referenda voting functions are Pareto efficient. In addition, the voters have the ability to lock their votes for an extended time, to signal the strength of their preferences. Arguably, a veto might not be Pareto efficient if there is $100\%$ consensus in a referendum. However, this is an extremely contrived case. For all intents and purposes governance \emph{is} Pareto efficient.
     \item \textbf{Confidentiality}: 
     \begin{itemize}
        \item \textbf{Secrecy}: Votes on Polkadot, whether it's in electing councillors, internal council votes, or voting in referenda, are not documented to be private. Therefore, secrecy is \emph{not} satisfied.
        \item \textbf{Pseudonymity}: Participants are not required to reveal their real-life identities to participate in the decision-making process.Therefore pseudonymity \emph{is} satisfied.
        \item \textbf{Coercion-resistance}: Since secrecy is \textit{not} satisfied, coercion-resistance is \emph{not} satisfied by definition.
    \end{itemize}
    \item \textbf{Verifiability}. Since the votes and final tally are all public, verifiability \emph{is} satisfied.
    \item \textbf{Accountability}. Voting in favour of a proposal requires funds to be locked in until the proposal is enacted. The documented rationale behind this is to hold voters responsible for a proposal that they vote for. Therefore, accountability \emph{is} satisfied.
    \item \textbf{Sustainability}: There are no explicit or direct rewards given for participation, but successful proposals requiring funds can access the treasury, after approval from the council. As mentioned in the main text, Polkadot has explicitly chosen against direct voting rewards. Sustainable development is only somewhat satisfied, as the current mechanism is still a bit informal.
    \item \textbf{Liveness}. The Polkadot governance mechanism is capable of taking in inputs of urgency (i.e. emergency referenda) and acting on it if deemed urgent by the council, all whilst being able to terminate within an amount of time proportional to the urgency.
    Therefore, liveness \emph{is} satisfied.
\end{itemize}

\subsection{Compound}\label{sec:compound}
Compound~\cite{compound_white_paper} is a protocol running on the Ethereum blockchain that establishes money markets. These are collections of Ethereum assets (e.g. Ether, ERC-20 stablecoins, coins like DAI or ERC-20 utility coins such as Augur) that users can supply and borrow. These assets have algorithmically defined interest rates, dependent on supply and demand, that users collect or pay when supplying and borrowing respectively. Users can borrow depending on the value of the underlying asset they have as collateral and repay at any rate they want, paying the accrued interest. This provides the ability to quickly switch between tokens in a trustless manner. 

Governance in Compound is fuelled by an ERC-20 compatible token called \href{https://etherscan.io/token/0xc00e94cb662c3520282e6f5717214004a7f26888}{COMP} \cite{comp_coinbase}. The maximum number of COMP tokens is capped at 10,000,000. About 4,200,000 of them are distributed to the community  at a rate of 2,312 per day. Of those, a fixed fraction of these tokens is allocated to every market on Compound, half of which goes to suppliers and the other half to borrowers and subsequently allocated proportionately within each group. Additionally, 2,400,000 tokens belong to the Compound Labs Inc. shareholders, 2,200,000 are allocated over 4 years to the Compound team (with an additional 320,000 reserver for future members) and finally 775,000 are reserved for the community.

Holders of COMP can delegate voting power and create \emph{government} proposals. COMP tokens can be delegated to other addresses at rate of 1 vote per token, or delegated to oneself for a direct vote. A government proposal can then be created by any address holding at least 25,000 COMP. On top of that, any address with 100 COMP can create an \emph{autonomous} proposal, which in turn can become a government proposal once that address receives 25,000 COMP or more in delegation. A government proposal is an executable piece of code, which could update some parameter (e.g. the rate at which COMP tokens are distributed), create a new money market or provide additional functionality to the Compound smart contracts. A single address cannot issue multiple proposals in parallel.

The governance process is controlled by two smart contracts: Governor Bravo and Timelock. Once a government proposal is created, it is put into a two day review period, followed by an election lasting 3 days. COMP holders can vote for or against the proposal, which passes if the majority was in favour \emph{and} it received more than 400,000 votes in total. After that, it is put in Timelock for a mandatory 2 day waiting period, before it is executed. This is a safety measure: if an issue is found while in Timelock, the proposer can cancel it (or the users can start reacting before its too late). At any point prior to execution, the creator of the proposal (or any address if the creator has fewer than 25,000 COMP) can cancel the process. In addition the \emph{Pause Guardian} (which is controlled by a community appointed multi-signature) can suspend the functionality of some Compound function (namely Mint, Borrow, Transfer, and Liquidate) allowing users only very benign actions such as closing their positions.

\begin{itemize}
    \item \textbf{Suffrage}.
    Since voting eligibility depends only on having COMP tokens, which can be exchanged and are initially distributed to addresses with assets on Compound, token-based suffrage \emph{is} satisfied. Some COMP tokens are distributed or reserved for members of the Compound team. Therefore, meritocratic suffrage is \emph{slightly} satisfied.
    
    \item \textbf{Pareto Efficiency}. Once a proposal enters the voting phase, the voters only have two options: yes or no. This is clearly Pareto efficient and aligned with their incentives. Things get more tricky once there are multiple incompatible options (e.g., values of a specific parameter). In this case the  proposals would have to be dealt with sequentially: the actual order could bias voters, which complicates their decisions and leaks information. Therefore, Pareto Efficiency is \emph{somewhat} satisfied (e.g., between two highly popular proposal, the slightly less popular one might win if it is up for election first and then the users might be less eager to implement another change).
    
    \item \textbf{Confidentiality}:
    \begin{itemize}
        \item \textbf{Secrecy and Coercion Resistance}: Every step of the governance process, such as proposing, voting or delegating is on-chain, by interacting with smart contracts on Ethereum. This done through possibly pseudonymous addresses and is public and unencrypted. Therefore, \emph{neither} property satisfied.
        \item \textbf{Pseudonymity}: Users participate using their Ethereum address, therefore pseudonymity \emph{is} satisfied.
    \end{itemize}
    \item \textbf{Verifiability}. Since the votes and final tally are all public, verifiability \emph{is} satisfied.
    
    \item \textbf{Accountability}. Once a proposal is executed, its creator and voters are completely independent from its future. Therefore, accountability is \textit{not} satisfied.
    \item \textbf{Sustainability}: 
    \begin{itemize}
    	\item \textbf{Sustainable Development}:
    		There is no mechanism to reward development efforts: the proposal 
    		should already be complete and executable. Therefore, sustainable 
    		development is \emph{not} satisfied.
    		\item \textbf{Sustainable Participation}: Although COMP tokens have an 
    		value and can be traded, there are no additional reward for voting or 
    		creating a government proposal. Therefore, sustainable participation is 
    		\emph{not} satisfied.
	\end{itemize}
    \item \textbf{Liveness}. The total time between creating a government proposal and voting for it takes 7 days, 2 of which are hard-coded into the Timelock. This is reasonable: in addition, if an exploit is found while in Timelock, the proposer can cancel it. Failing to do so, the users of Compound have some time to either move their assets or fork. This window for immediate action is typically only open right after a vote, however the Pause Guardian ensures that an `emergency shutdown' feature is always available. Therefore, liveness \emph{is} satisfied.
\end{itemize}

\subsection{Maker DAO}
Maker DAO~\cite{makerdao_white_paper} is a decentralized organization running on Ethereum and based on the Maker Protocol. It employs a two-token system, using Dai and MKR, both of which are ERC-20 compatible. The first, DAI, is a collateral-backed stablecoin which is soft-pegged to the U.S. dollar and is collateralized by a \emph{mix} of other cryptocurrencies. The second, MKR, is a governance token is used by stakeholders to maintain the system and manage Dai. However, in addition to the previous governance token models, MKR, which is \emph{not} a stablecoin, is also used to control the price of Dai, by creating favourable exchange rates between the two coins, depending on Dai supply and demand. In particular, 1,000,000 MKR were originally minted. The total supply is then kept as close to this number as possible, by burning or minting new tokens in exchange for Dai.  

The governance model employed~\cite{makerdao_community} combines some of the features of Compound (such as on-chain voting for some issues, executable proposals and a mandatory waiting period) and some off-chain features of Uniswap (such as forum discussions). Note that the two components are \emph{not} officially coupled. The off-chain component takes place at the Maker DAO \href{https://forum.makerdao.com/}{forum}, which is public. In addition to usual forum posts, users can (and are encouraged to) create a \emph{Forum Signal Thread}. The purpose is to get community feedback on some issue, possible on-chain proposals or generally any potential improvement to Maker DAO. At the end, the Forum Signal Thread is followed by a poll, where users vote pseudonymously.  Every user has \emph{one} vote, irrespective on the amount of MKR they may have. The intended function is that the discussion and poll results will inform the choices of an upcoming \emph{on-chain} governance action.

There are two on-chain processes facilitated by smart contracts: \emph{Governance Polls} and \emph{Executive Votes}. The aim of Governance Polls is to ratify Forum Signal Threads, formally gauge consensus about important topics and select one of many alternative designs before an Executive Vote. The Governance Poll could contain multiple options and holders of MKR vote using instant-runoff. Governance Polls usually stay open for 3 to 7 days. The results of Governance Polls can then be turned into Executive Votes, although both processes could be initiated by any Ethereum address at any point. However, only Governance Facilitators can link specific Governance Polls and Executive Votes in the official forum.

The Executive Vote is the only way to enact changes on the smart contracts supporting of Maker DAO. Indeed, an Executive Vote should contain instructions to amend their code with the proposed set of changes. Executive Votes are selected via \emph{continuous} approval voting, typically without having a fixed voting window. Specifically, holders of MKR can change their vote at any time and the Executive Vote with the highest approval would win. However, once an Executive Vote that was implemented loses to another one, it is deactivated and the only way to revert to the previous status is through a new vote. Once a new Executive Vote wins, the Governance Security Module imposes a 24 hour waiting period, during which the vote can be reversed.

Maker DAO also makes use of Emergency Shutdown. At any point if a total of 50,000 MKR are deposited into the Emergency Shutdown Module, an Emergency Shutdown is triggered. These coins are immediately burned and the Maker Protocol is shut down. Then, collateral supporting Dai (as well as the coins themselves) are returned to their owners. For various reasons, Dai takes lower priority than collateral and could be exchanged for less than 1\$ per Dai.

\begin{itemize}
    \item \textbf{Suffrage}.
        Since voting eligibility is only guaranteed to MKR token holders, token-based suffrage \emph{is} satisfied.

    \item \textbf{Pareto Efficiency}. For Executive Votes, the voters only have two options: to vote yes or no. Even though these do not have to follow Governance Polls, the ranked-choice, instant runoff voting mechanism used there gives the voters the option to choose between multiple alternatives, avoiding the possibility of a sequential vote (e.g., as could happen in Compound). Therefore, Pareto Efficiency is \emph{mostly} satisfied.
    
    \item \textbf{Confidentiality}: 
    \begin{itemize}
        \item \textbf{Secrecy}: Every step of the governance process, such as proposing, voting or delegating is on-chain, by interacting with smart contracts on Ethereum. This done through possibly pseudonymous addresses and is public and unencrypted. Therefore secrecy is \emph{not} satisfied.
        \item \textbf{Pseudonymity}: Users participate using their Ethereum address. Therefore, pseudonymity \emph{is} satisfied.
        \item \textbf{Coercion-resistance}: Since secrecy is not satisfied, coercion-resistance is \emph{not} satisfied by definition.
    \end{itemize}   
    \item \textbf{Verifiability}. Since the votes and final tally are all public, verifiability \emph{is} satisfied.
    \item \textbf{Accountability}. As with Compound, once a proposal is executed, its creator and voters are completely independent from its future. Therefore, accountability is \textit{not} satisfied.
    \item \textbf{Sustainability}: 
    \begin{itemize}
    	\item \textbf{Sustainable Development}:
    		There is no mechanism to reward development efforts: the proposal 
    		should already be complete and executable. Therefore, sustainable 
    		development is \emph{not} satisfied.
    		\item \textbf{Sustainable Participation}: MKR tokens are crucial for the 
    		economy and function of Maker DAO. However, the extra energy spent on 
    		deciding what to vote on is not explicitly compensated. Therefore, 
    		sustainable participation is \emph{not} satisfied.
    	\end{itemize}
    \item \textbf{Liveness}. An Executive Vote can be implemented in 24 hours, once it receives enough votes. This gives both the ability to quickly prevent a bad proposal and relatively quickly enact a better one. In addition, there is also an Emergency Shutdown functionality. Therefore, liveness \emph{is} satisfied.
\end{itemize}

\subsection{\textbf{Project Catalyst}}
\label{sec:catalyst}
Project Catalyst \cite{catalyst_community} is the on-chain governance system used by the Cardano blockchain. The role of Project Catalyst is to provide a mechanism through which users can collectively decide how Cardano's treasury funds should be allocated. 

Governance in Project Catalyst occurs in 12 week intervals, called \emph{Funds}. There are 4 primary types of agents participating: proposers, voters, Community Advisors (CA's) and Veteran Community Advisors (vCA's). Additionally, people can participate by referring projects to be funded and designing challenges that need to be addressed. Finally, the \emph{Catalyst Circle} \cite{catalyst_circle} is a small group of representatives of all types of agents involved, tasked with monitoring the current state and developing future plans for Project Catalyst. The Circle is currently not elected, but an election mechanism is discussed for future iterations. At the beginning of each fund a set of challenges is issued, either by users of Cardano or the Project Catalyst team. Then, the proposers offer proposals, which may, but are not required to, address a specific challenge. The proposals should contain a detailed set of goals, along with a specific plan to achieve them and a required budget. Then, the community advisors write reviews for any proposal they chose to, focusing on impact, implementability and auditability. These reviews are then reviewed again by the veteran community advisors and are assigned a grade that can be `Excellent', `Good' or `Filtered Out', the last reserved for particularly uninformative reviews. Having all this information, the voters can vote `Yes', `No' or `Abstain' for as many proposals as they want. Each vote has weight proportional to the users stake in ADA, which is the currency used by Cardano. Project Catalyst implements \textit{fuzzy threshold voting}~\cite{zhang2019treasury}. Voters express a `Yes', `No' or `Abstain' opinion for each proposal. A proposal passes if the number of `Yes' votes minus the number of `No' votes is at least $5\%$ of the total votes it received. The winning proposals are awarded their funds in the order of the margin by which they are passing, until either the entire budget is allocated or no more passing proposals exist. If a proposal has passed the voting threshold but insufficient funds remain to pay the full amount requested, it will not receive partial funding. Instead, any smaller proposals which have also passed the threshold that will fit in the budget will be funded, even if they have lower net approval than the larger proposal.

All agents involved in Project Catalyst are rewarded in some capacity. At every Fund each reward pool corresponds to a set percentage of the total. As a concrete example we will examine \href{https://www.reddit.com/r/cardano/comments/qwngps/project_catalyst_fund7_voter_registration/}{Fund7}, which had total budget of \$8,000,000 in ADA. This amount was further broken down as follows:
\begin{itemize}
    \item $80\% \rightarrow \$6,400,000$ for funding proposals
    \item $13\% \rightarrow \$1,040,000$ for voting rewards.
    \item $4\% \rightarrow \$320,000$ for community advisors
    \item $1\% \rightarrow \$80,000$ for veteran Community Advisors.
    \item $1\% \rightarrow \$80,000$ for referral rewards.
    \item $1\% \rightarrow \$80,000$ for challenge teams rewards.
\end{itemize}
Any user with more than $500$ ADA can become a voter. This is measured by a snapshot of the stake distribution taken before the election, but the funds are not locked. Each voter receives voter rewards proportional to their stake. Community advisors receive rewards relative to the quality of the reviews, but also depending on how many other reviews were written for the proposals they reviewed. An `Excellent' review provides 3 times the reward of a `Good' review and each proposal has rewards for $2$ `Excellent' and $3$ `Good' reviews. If these rewards are not enough to cover the reviews, a lottery is used. Veteran community advisors are rewarded equally, provided they reviewed a minimum number of reviews. Proposers are not rewarded explicitly, but can manage the funds received by their proposal and have to periodically submit progress reports to the community. The performance of community advisors and veteran community advisors is recorded, but there is no currently defined on-chain mechanism for a voter to become either of those. The promotion from voter (or proposer) to community advisor to veteran is centralized.

\begin{itemize}
    \item \textbf{Suffrage}.
        Since voting eligibility depends only on having at least $500$ ADA, token-based suffrage \emph{is} satisfied. There are no guaranteed voting rights based on previous positive contributions. However, community advisors and veteran community advisors can affect the outcome of the votes through their reviews. Meritocratic suffrage is \emph{slightly} satisfied.
        
        \item \textbf{Pareto Efficiency}. As noted in the main text evaluation, Pareto Efficiency is only \emph{somewhat} satisfied.
        
     \item \textbf{Confidentiality}:
     \begin{itemize}
        \item \textbf{Secrecy}: Everyone participates in Project Catalyst using their wallet address. Proposers, community advisors and veteran community advisors participate publicly. Voters submit \emph{encrypted} ballots (padded with some randomness), using the public key issued by a committee. Then, these votes are tallied and the result is decrypted by the committee, if a majority of its members agrees. Furthermore, if the wallet address is linked to a real identity, the only information available is that this particular person voted, but the actual vote is still secret.
        Therefore the vote is \emph{mostly} secret. 
        \item \textbf{Pseudonymity}: Voters participate with their wallet address, therefore pseudonymity \emph{is} satisfied.
        \item \textbf{Coercion-resistance}: The system is somewhat coercion resistant. The ballot itself cannot be decrypted by the voter. Additionally, if the random padding is not kept, it is impossible even for the voter to convince anyone of the way they voted.
    \end{itemize}
    \item \textbf{Verifiability}. The result of the vote can be independently verified. In addition, as long as a voter saved the random padding, they can verify that their particular vote was counted. Without the padding this is impossible, as the votes \emph{cannot} be decrypted. As such, verifiability is only \emph{mostly} satisfied.

    \item \textbf{Accountability}. There are no explicit, on or off-chain, penalties. Proposers need to submit periodic progress reports about their projects to keep receiving funding. Similarly, community advisors and veteran community advisors can be penalized for poor reviews or absence. As these are either centralized or community-driven without clearly described mechanisms, accountability is mostly \emph{not} satisfied.
    \item \textbf{Sustainability}:
    \begin{itemize}
    	\item \textbf{Sustainable Development}: Although there is no explicit 
    	incentive or reward given to the proposing group or individual, it is the 
    	responsibility of the proposer to request the amount which represents the 
    	value of their work. Therefore, sustainable development \textit{is} 
    	satisfied.
    	\item \textbf{Sustainable Participation}:
    	Since all parties are rewarded for participating in the governance process 
    	and to an extent receive larger rewards for additional effort (e.g. 
    	community advisors and review quality), sustainable participation 
    	\emph{is} satisfied. 
   	\end{itemize}
    \item \textbf{Liveness}. Project Catalyst is primarily used for allocating treasury funds and each Fund follows a 12 week timeline. As such, liveness is \emph{not} satisfied: even though the funds can be released in accordance with each proposal's progress, there is no direct mechanism to take urgent action. However, liveness is arguably not required for its purposes.
\end{itemize}

\subsection{\textbf{Dash}}
\label{sec:dash}
Like Bitcoin, Dash \cite{dash_docs} uses a proof-of-work consensus mechanism. However, Dash's approach to governance takes a formal, on-chain form. The Dash Governance System (DGS) uses a `budget and masternode voting system' to govern and fund the underlying blockchain’s development and maintenance. Masternodes are nodes that can place at least a $1,000$ DASH, the platform's native token, as a collateral to participate in the consensus protocol and governance process. Each masternode has a single, public, approval vote expressing which improvement proposals the masternode approves of. In each voting cycle (which is roughly a month long), project proposals are submitted and then voted on. Even though anyone can submit a proposal, doing so comes at a cost of 5 DASH to ensure that only serious proposals are voted on.

The DGS implements a system very similar to Project Catalyst with one difference: A proposal is eligible for funding if the number of `Yes' votes minus the number of `No' votes is at least $10\%$ of the \emph{total} masternode count. Additionally, if there are two proposals with the same approval, then the one with a larger proposal transaction hash is ranked higher. The treasury is funded through various channels. When new blocks are mined, $45\%$ of the block reward is reserved for the miner, $10\%$ for the budget and $45\%$ for the masternodes’ reward. 
We now perform evaluations of the DGS against the properties developed in Section \ref{section:properties}.

\begin{itemize}
     \item \textbf{Confidentiality}: 
     \begin{itemize}
         \item \textbf{Secrecy}: Since the masternodes vote publicly, the DGS does \textit{not} satisfy secrecy.
        \item \textbf{Pseudonymity}: Masternodes are not required to reveal their real-life identities to participate in the governance process; therefore pseudonymity is satisfied.
        \item \textbf{Coercion-resistance}: Since masternode votes are tied to their chosen pseudo-identities, coercion-resistance is \textit{not} satisfied.
    \end{itemize}   
    \item \textbf{Verifiability}. Since the votes and final tally are all public, verifiability is, by definition, satisfied.
    \item \textbf{Pareto Efficiency}. As with Project Catalyst, Pareto Efficiency is only \emph{somewhat} satisfied. 
    \item \textbf{Accountability}. Although masternodes are required to lock $1,000$ DASH to vote, if a group of masternodes vote in a malicious proposal, they will face no negative consequences and will be able to unlock their funds before the malicious proposal is enacted. Therefore, accountability is \textit{not} satisfied.
    \item \textbf{Sustainability}: 
    \begin{itemize}
    	\item \textbf{Sustainable Development}: Although there is no explicit 
    	incentive or reward given to the proposing group or individual, it is the 
    	responsibility of the proposer to request the amount which represents the 
    	value of their work. Therefore, sustainable development \textit{is} 
    	satisfied.
    	\item \textbf{Sustainable Participation}: Masternodes are rewarded with 
    	part of the block reward for their participation in the consensus and 
    	governance process. Therefore, sustainable participation is satisfied.
   	\end{itemize}
    \item \textbf{Liveness}. Given the lack of flexibility of the on-chain governance model, the DGS is incapable of taking inputs of urgency and responding to them in accordance to the severity of the issue. Although a Github issue or a pull request could be initiated without going through the formal on-chain route, it is still not the officially defined way by which the system processes inputs. Therefore, liveness is \textit{not} satisfied.
    \item \textbf{Suffrage}.
    Since voting eligibility depends only on having at least $1,000$ DASH, token-based suffrage is satisfied.
    \end{itemize}

\subsection{Decred}
Decred is a hybrid proof-of-work and proof-of-stake system that is mostly on-chain governed \cite{decred_documentation}. Such a hybrid implementation results in three main types of stakeholders: miners, voters and regular users. All three participate pseudo-anonymously. To have decision-making powers (in governance and block-validation), participants need to have `tickets', which are bought or acquired through time-locking DCR (the native token of the platform). We will not go through the details of a ticket lifecycle, but the process is thoroughly outlined in \cite{decred_documentation}. Each block contains 5 pseudo-randomly sampled tickets (i.e. 5 votes).

Proposals can be handled either by an on-chain or off-chain procedure. Specifically, proposals regarding high level issues or that require funds from the Decred Treasury are handled off-chain. They first appear in \href{https://proposals.decred.org/}{Politeia}, the system's deliberation platform, to be discussed throughout the community. Administrators of the platform can flag spam proposals or comments. When a proposal owner decides to put their proposal for a vote, the administrators can then trigger the start of off-chain voting. A snapshot of the currently bought tickets takes place 256 blocks before the start of voting. Then, the ticket-voting interval of 2,016 blocks (approximately 1 week) formally begins, which means 10,080 pseudo-randomly sampled tickets have the opportunity to vote. Voting on Politeia is not recorded on chain, but it is still backed by cryptographic techniques which prevent Sybil attacks and unfair censorship. When the ticket-voting period ends, the proposal is formally approved or rejected. There is a quorum requirement for a vote to be considered valid: 20\% of the eligible tickets must vote `Yes’ or `No’. The threshold for a proposal to be approved is 60\% `Yes’ votes. When a proposal with a budget and deliverables is approved, work can begin. The proposal owner can submit claims against the budget as deliverables are completed.

The on-chain governance is performed through Decred Change Proposal (DCP) \cite{decred_change_proposals}, focusing on updating the consensus mechanism. With a DCP, the proposed node software must be developed and released. The new code will lie dormant until the change has been voted upon and accepted by the proof-of-stake voters. Each voting interval lasts for 8,064 blocks, which makes the maximum number of votes 40,320. A ticket can vote to accept the rule change, to reject it or to abstain (the default choice). Every vote has a quorum requirement of 10\%. This means that at least 10\% of all votes cast must be non-abstain for the result to be considered valid. If all non-abstaining votes fail to meet a 75\% Yes or No majority threshold, the agenda vote remains active for next voting period. If 75\% of all non-abstaining votes accept the proposal, the agenda is considered locked in and the consensus changes will activate 8,064 blocks (4 weeks) after the vote passed. If 75\% of all non-abstaining votes reject the proposal, the agenda fails and the consensus changes will never activate. If an agenda reaches its expiration before ever reaching a 75\% majority vote, the agenda expires and the consensus changes will never activate. After a ticket has voted, missed, or expired, the funds cannot be released for another 256 blocks.

If the quorum requirement is met, and more than 75\% of the votes are in favour of activating the new consensus rules, then a `lock-in' period begins of 8,064 blocks. During this period, all participants in the Decred network must upgrade their software to the latest version. All full nodes participating in the network will automatically activate the new rules on the first block after this period, so any nodes still running the old software will no longer be able to participate. Throughout the process, it is possible to verify the voting preference of a ticket. 

With this brief overview in mind, we can now perform the evaluation of Decred's governance system against our properties. 

\begin{itemize}
    \item \textbf{Suffrage}. Since voting eligibility only depends on buying proof-of-stake tickets, token-based suffrage \emph{is} satisfied.

    \item \textbf{Pareto Efficiency}. If a proposal vote occurs with a quorum of less than 10\%, the proposal will not pass, even when it receives one or more approval votes. Furthermore, given the role of Politeia, it is unlikely that a truly controversial proposal will pass. Therefore, the most likely `suboptimal' outcome is not selecting any proposal, when one might have had some support. Therefore, Pareto efficiency is \textit{somewhat} satisfied.

     \item \textbf{Confidentiality}:
     \begin{itemize}
        \item \textbf{Secrecy}: There are no explicit secrecy guarantees in the voting process. Therefore, secrecy is \textit{not} satisfied.
        \item \textbf{Pseudonymity}: Participants (miners, voters, and regular users) are not required to reveal their real-life identities to participate in the decision-making process. Therefore pseudonymity \emph{is} satisfied.
        \item \textbf{Coercion-resistance}: Since secrecy is not satisfied, coercion-resistance is \emph{not} satisfied by definition.
    \end{itemize}
    \item \textbf{Verifiability}. Since the votes and final tally are all public, verifiability \emph{is} satisfied.
    \item \textbf{Accountability}. Although funds from the ticket cannot be released until 256 blocks after voting, the changes to the consensus rules are not applied until after 8,064 blocks. This implies that if a voter or a group of voters voted in a malicious proposal, they can withdraw their locked funds before the proposal is enacted. Therefore, accountability is \textit{not} satisfied.
    \item \textbf{Sustainability}: 
    \begin{itemize}
    	\item \textbf{Sustainable Development}: Although there is no explicit 
    	incentive or reward given to the proposing group or individual, it is the 
    	responsibility of the proposer to request the amount which represents the 
    	value of their work. Therefore, sustainable development \textit{is} 
    	satisfied.
    	\item \textbf{Sustainable Participation}: Although voters can gain rewards 
    	from their tickets via validating blocks as part of the consensus 
    	protocol~\cite{decred_documentation}, there are no explicit additional 
    	incentives for voting (or participating in the governance process). 
    	Therefore, sustainable participation is \textit{not} satisfied.
   	\end{itemize}
    \item \textbf{Liveness}. Given the lack of flexibility of the on-chain governance model, it is incapable of taking inputs of urgency and responding to them in accordance to the severity of the issue. Therefore, liveness is \textit{not} satisfied.
\end{itemize}

\end{document}